\documentclass[12pt]{article}
\usepackage{a4wide,graphicx,psfrag,rotating,epsfig,color}
\usepackage{amsmath}
\allowdisplaybreaks[1]

\makeatletter

\def\paragraph{\@startsection{paragraph}{4}{\z@}{+2.00ex plus
 +1ex minus +.2ex}{1.5ex plus .2ex}{\it\normalsize}}

\def\section{\@startsection {section}{1}{\z@}{+3.0ex plus +1ex minus
  +.2ex}{2.3ex plus .2ex}{\normalsize\bf\boldmath}}
\def\subsection{\@startsection{subsection}{2}{\z@}{+2.5ex plus +1ex
minus +.2ex}{1.5ex plus .2ex}{\normalsize\bf\boldmath}}
\def\subsubsection{\@startsection{subsubsection}{3}{\z@}{+3.25ex plus
 +1ex minus +.2ex}{1.5ex plus .2ex}{\normalsize\it}}

\expandafter\ifx\csname mathrm\endcsname\relax\def\mathrm#1{{\rm #1}}\fi

\newcounter{saveeqn}

\@addtoreset{equation}{section}

\newcount\@tempcntc
\def\@citex[#1]#2{\if@filesw\immediate\write\@auxout{\string\citation{#2}}\fi
  \@tempcnta\z@\@tempcntb\m@ne\def\@citea{}\@cite{\@for\@citeb:=#2\do
    {\@ifundefined
       {b@\@citeb}{\@citeo\@tempcntb\m@ne\@citea
        \def\@citea{,\penalty\@m\ }{\bf ?}\@warning
       {Citation `\@citeb' on page \thepage \space undefined}}%
    {\setbox\z@\hbox{\global\@tempcntc0\csname
b@\@citeb\endcsname\relax}%
     \ifnum\@tempcntc=\z@ \@citeo\@tempcntb\m@ne
       \@citea\def\@citea{,\penalty\@m}
       \hbox{\csname b@\@citeb\endcsname}%
     \else
      \advance\@tempcntb\@ne
      \ifnum\@tempcntb=\@tempcntc
      \else\advance\@tempcntb\m@ne\@citeo
      \@tempcnta\@tempcntc\@tempcntb\@tempcntc\fi\fi}}\@citeo}{#1}}

\def\@citeo{\ifnum\@tempcnta>\@tempcntb\else\@citea
  \def\@citea{,\penalty\@m}%
  \ifnum\@tempcnta=\@tempcntb\the\@tempcnta\else
   {\advance\@tempcnta\@ne\ifnum\@tempcnta=\@tempcntb \else
\def\@citea{--}\fi
    \advance\@tempcnta\m@ne\the\@tempcnta\@citea\the\@tempcntb}\fi\fi}

\def\nl{\nonumber\\}

\newcommand{\lsim}
{\mathrel{\raisebox{-.3em}{$\stackrel{\displaystyle <}{\sim}$}}}
\newcommand{\gsim}
{\mathrel{\raisebox{-.3em}{$\stackrel{\displaystyle >}{\sim}$}}}
\def\asymp#1%
{\mathrel{\raisebox{-.4em}{$\widetilde{\scriptstyle #1}$}}}

\def\Nequal#1%
{\mathrel{\raisebox{-.5em}{$\stackrel{=}{\scriptstyle\rm#1}$}}}
\newcommand{\dsl}[1]{\not \hspace{-0.7mm}#1}
\def\dsl{\mathpalette\make@slash}
\def\make@slash#1#2{\setbox\z@\hbox{$#1#2$}%
  \hbox to 0pt{\hss$#1/$\hss\kern-\wd0}\box0}
\def\eqref#1{\stackrel{\mbox{\scriptsize{(\ref{#1})}}}{=}}

\makeatother

\def\beq{\begin{equation}}
\def\eeq{\end{equation}}
\def\beqar{\begin{eqnarray}}
\def\eeqar{\end{eqnarray}}
\def\barr#1{\begin{array}{#1}}
\def\earr{\end{array}}
\def\bfi{\begin{figure}}
\def\efi{\end{figure}}
\def\btab{\begin{table}}
\def\etab{\end{table}}
\def\bce{\begin{center}}
\def\ece{\end{center}}

\def\text{\textstyle}

\def\al{\alpha}
\def\Ga{\Gamma}
\def\ga{\gamma}
\def\de{\delta}
\def\De{\Delta}

\def\la{\lambda}

\def\si{\sigma}

\def\refeq#1{\mbox{(\ref{#1})}}

\def\reffi#1{\mbox{Figure~\ref{#1}}}

\def\refta#1{\mbox{Table~\ref{#1}}}

\def\refse#1{\mbox{Section~\ref{#1}}}

\def\refapp#1{\mbox{App.~\ref{#1}}}
\def\citere#1{\mbox{Ref.~\cite{#1}}}
\def\citeres#1{\mbox{Refs.~\cite{#1}}}

\newcommand{\TeV}{\unskip\,\mathrm{TeV}}
\newcommand{\GeV}{\unskip\,\mathrm{GeV}}

\newcommand{\fb}{\unskip\,\mathrm{fb}}

\newcommand{\ri}{{\mathrm{i}}}

\newcommand{\rw}{\mathswitchr w}

\newcommand{\Oa}{\mathswitch{{\cal{O}}(\alpha)}}

\def\mathswitchr#1{\relax\ifmmode{\mathrm{#1}}\else$\mathrm{#1}$\fi}

\newcommand{\CM}{\mathswitchr {CM}}

\newcommand{\PW}{\mathswitchr W}
\newcommand{\PZ}{\mathswitchr Z}

\newcommand{\PH}{\mathswitchr H}
\newcommand{\Pe}{\mathswitchr e}
\newcommand{\Pp}{\mathswitchr p}

\newcommand{\Pd}{\mathswitchr d}

\newcommand{\Pu}{\mathswitchr u}

\newcommand{\Ps}{\mathswitchr s}

\newcommand{\Pc}{\mathswitchr c}

\newcommand{\Pb}{\mathswitchr b}

\newcommand{\Pt}{\mathswitchr t}

\newcommand{\Pep}{\mathswitchr {e^+}}
\newcommand{\Pem}{\mathswitchr {e^-}}
\newcommand{\PWp}{\mathswitchr {W^+}}
\newcommand{\PWm}{\mathswitchr {W^-}}

\def\mathswitch#1{\relax\ifmmode#1\else$#1$\fi}

\newcommand{\MW}{\mathswitch {M_\PW}}
\newcommand{\MA}{\mathswitch {m_\ga}}
\newcommand{\MZ}{\mathswitch {M_\PZ}}
\newcommand{\MH}{\mathswitch {M_\PH}}
\newcommand{\Me}{\mathswitch {m_\Pe}}

\newcommand{\Md}{\mathswitch {m_\Pd}}
\newcommand{\Mu}{\mathswitch {m_\Pu}}
\newcommand{\Ms}{\mathswitch {m_\Ps}}
\newcommand{\Mc}{\mathswitch {m_\Pc}}
\newcommand{\Mb}{\mathswitch {m_\Pb}}
\newcommand{\Mt}{\mathswitch {m_\Pt}}

\newcommand{\GW}{\Gamma_{\PW}}
\newcommand{\GZ}{\Gamma_{\PZ}}

\newcommand{\sw}{\mathswitch {s_\rw}}
\newcommand{\cw}{\mathswitch {c_\rw}}

\newcommand{\Qu}{\mathswitch {Q_\Pu}}
\newcommand{\Qd}{\mathswitch {Q_\Pd}}

\newcommand{\GF}{\mathswitch {G_\mu}}

\def\solid{\raise.9mm\hbox{\protect\rule{1.1cm}{.2mm}}}
\def\dash{\raise.9mm\hbox{\protect\rule{2mm}{.2mm}}\hspace*{1mm}}
\def\ie{i.e.\ }
\def\eg{e.g.\ }
\def\cf{cf.\ }

\newcommand{\virt}{{\mathrm{virt}}}

\def\Re{\mathop{\mathrm{Re}}\nolimits}

\newcommand{\arc}{\mathop{\mathrm{arc}}\nolimits}

\newcommand{\dels}{\de_\mathrm{s}}
\newcommand{\delc}{\de_\mathrm{c}}
\newcommand{\IRsafe}{\MSbar}
\newcommand{\singf}{\mathrm{sing}}
\newcommand{\Ppo}{\Pp}
\newcommand{\Ppt}{\Pp}

\def\draftdate{\relax}
\def\mda{\relax}
\def\mua{\relax}
\def\mla{\relax}
\def\draft{
\def\thtystars{******************************}
\def\sixtystars{\thtystars\thtystars}
\typeout{}
\typeout{\sixtystars**}
\typeout{* Draft mode!
         For final version remove \protect\draft\space in source file *}
\typeout{\sixtystars**}
\typeout{}
\def\draftdate{\today}
\def\mua{\marginpar[\boldmath\hfil$\uparrow$]%
                   {\boldmath$\uparrow$\hfil}%
                    \typeout{marginpar: $\uparrow$}\ignorespaces}
\def\mda{\marginpar[\boldmath\hfil$\downarrow$]%
                   {\boldmath$\downarrow$\hfil}%
                    \typeout{marginpar: $\downarrow$}\ignorespaces}
\def\mla{\marginpar[\boldmath\hfil$\rightarrow$]%
                   {\boldmath$\leftarrow $\hfil}%
                    \typeout{marginpar: $\leftrightarrow$}\ignorespaces}
\def\Mua{\marginpar[\boldmath\hfil$\Uparrow$]%
                   {\boldmath$\Uparrow$\hfil}%
                    \typeout{marginpar: $\uparrow$}\ignorespaces}
\def\Mda{\marginpar[\boldmath\hfil$\Downarrow$]%
                   {\boldmath$\Downarrow$\hfil}%
                    \typeout{marginpar: $\downarrow$}\ignorespaces}
\def\Mla{\marginpar[\boldmath\hfil$\Rightarrow$]%
                   {\boldmath$\Leftarrow $\hfil}%
                    \typeout{marginpar: $\leftrightarrow$}\ignorespaces}
\overfullrule 5pt
\oddsidemargin -15mm
\marginparwidth 29mm
}

\def\stars{\strut\leaders\hbox{*}\hfill\strut}
\def\starline{\hfil\strut\hfil\hbox to \textwidth {\stars}\hfil}

\newcommand{\pr}[2]{\langle #1 #2 \rangle}
\newcommand{\cpr}[2]{\langle #1 #2 \rangle^{\star}}
\newcommand{\spr}[4]{\langle #1\vert #2 + #3 \vert #4 \rangle}
\newcommand{\Abs}[1]{\mathop{\left\vert #1 \right\vert^2}}

\newcommand{\be}{\begin{equation}}
\newcommand{\ee}{\end{equation}}
\newcommand{\bea}{\begin{eqnarray}}
\newcommand{\eea}{\end{eqnarray}}

\newcommand{\MSbar}{\overline{\rm MS}}
\newcommand{\Pole}{{\tt Pole}}
\newcommand{\FA}{{\tt FeynArts}}
\newcommand{\FC}{{\tt FormCalc}}

\begin{document}
\bibliographystyle{h-elsevier3}

\voffset -1cm   % for hep-ph submission

\thispagestyle{empty}
\def\thefootnote{\fnsymbol{footnote}}
\setcounter{footnote}{1}
\null
\draftdate\strut\hfill PSI-PR-05-09\\
\strut\hfill ZU-TH 18/05 \\
\strut\hfill DFTT 30/05 %\\
\vfill
\begin{center}
{\Large \bf\boldmath
Electroweak corrections to $\PW\gamma$ and $\PZ\gamma$ production\\  at the LHC
\par} \vskip 1.5em
\vspace{.5cm}
{\large
{\sc E. Accomando$^1$, A.\ Denner$^2$, C.\ Meier$^{2,3}$}} \\[1cm]
$^1$ {\it Dipartimento di Fisica Teorica, Universit\`a di Torino,\\
Via P. Giuria 1, 10125 Torino, Italy} \\[0.5cm]
$^2$ {\it Paul Scherrer Institut, W\"urenlingen und Villigen\\
CH-5232 Villigen PSI, Switzerland} \\[0.5cm]
$^3$ {\it Institute for Theoretical Physics\\ University of Z\"urich, CH-8057 
Z\"urich, Switzerland}
\par \vskip 1em
\end{center}\par
\vfill \vskip 1cm {\bf Abstract:} \par 
We have calculated the electroweak {\rm ${\cal O}$}($\alpha$)
corrections to the processes $\Pp\Pp\rightarrow \PZ\gamma + X
\rightarrow \nu_l\bar{\nu}_l \gamma + X$, $\Pp\Pp\rightarrow \PZ\gamma
+ X \rightarrow l\bar{l} \gamma + X$, and $\Pp\Pp\rightarrow \PW\gamma
+ X \rightarrow \nu_l\bar{l} \gamma + X$ at the LHC, with $l=\Pe,\mu$
and $\nu_l=\nu_{\Pe},\nu_\mu,\nu_\tau$. The virtual corrections are
evaluated in leading-pole approximation, whereas the real corrections
are taken into account exactly. These corrections are implemented into
a Monte Carlo generator which includes both phase-space slicing and
subtraction to deal with soft and collinear singularities. We present
numerical results for total cross sections as well as for
experimentally interesting distributions. Applying typical LHC cuts,
the electroweak corrections are of the order of $-5\%$ for the total
cross sections and exceed $-20\%$ for observables dominated by high
centre-of-mass energies of the partonic processes.
\par
\vskip 1cm
\noindent
September 2005
\null
\setcounter{page}{0}
\clearpage
\def\thefootnote{\arabic{footnote}}
\setcounter{footnote}{0}

\section{Introduction}
\label{Introduction}
One of the primary goals of future colliders is to look for effects of
theories beyond the Standard Model (SM). In general, there are two
possibilities for new physics. Either the novel phenomena manifest
themselves at energy scales probed directly at forthcoming experiments
or at much larger scales.  In the first case one could admire a
spectacular scenario, characterized by the appearance of many new
particles or resonances. In the latter case, the situation appears
less favourable, but new-physics effects could still be detected in an
indirect way.  They can, for instance, influence physical observables
at low energy by modifying the structure of the gauge-boson
self-interactions.  Such modifications can be parametrized in terms of
anomalous couplings in the Yang--Mills vertices.  While the
gauge-boson couplings to the fermions have been measured at LEP2 and
Tevatron to an accuracy of $0.1$--$1\%$ \cite{ewwg}, triple and
quartic gauge-boson couplings have been determined to much lower
accuracy \cite{ewwg,gaugecouplep}.  Hence, the possibility of the
existence of anomalous couplings in the electroweak gauge sector
cannot be ruled out yet.

Vector-boson pair-production processes turn out to be especially
suited for testing the non-Abelian structure of the SM. In the last
decade, their potential has been extensively exploited at LEP2 and
Tevatron.  For these processes, the effect of anomalous couplings is
expected to rise strongly with increasing invariant mass of the
gauge-boson pair,
%$M_{\PV\PV^\prime}$ ($\PV,\PV^\prime = \PW,\PZ,\gamma$), 
since non-standard terms lead in general to unitarity violation for
longitudinal gauge bosons. At future colliders, it will thus be
worthwhile to analyse the large di-boson invariant-mass region.
Another interesting property of the vector-boson pair-production
processes is the so-called radiation zero, which characterizes the SM
partonic amplitudes of $\PW\gamma$ and $\PW\PZ$ production.  Owing to
gauge cancellations, for a given value of the scattering angle in the
di-boson rest frame, the amplitude for $\PW\gamma$ production vanishes
exactly and the one for $\PW\PZ$ approximately. For hadronic
processes, the resulting effect is the appearance of a dip in the
angular distributions (see \eg \citere{Baurradzero} and references
therein).  Any deviation from the SM gauge structure tends to modify
this peculiar signature, thus providing a clean possibility to look
for new physics.  Present experiments are not sensitive enough to see
the dip \cite{Tevatronradzero1,Tevatronradzero2}, but the LHC
offers the possibility to observe it for the first time.

Finally, the production of vector-boson pairs constitutes an important
background in the search for new particles.  For example, $\gamma\PZ$
production with the $\PZ$~boson decaying into a pair of neutrinos
gives rise to a photon-plus-missing-energy signature. The same signal
is expected for processes where the photon is produced in association
with one or more heavy particles which escape the detector, either
because they are weakly interacting or because they dissociate into
invisible decay products. Many extensions of the SM predict the
existence of such processes.  For example, the {\tt ADD} model
\cite{stringmodel1} and related models
\cite{stringmodel2,stringmodel3}, where gravity becomes strong at
$\TeV$ energies, predict the possibility to observe the production of
a photon along with a graviton. Higher-dimensional gravitons appear as
massive spin-2 neutral particles, which cannot be directly observed in
a detector.  Hence, graviton radiation leads to a missing-energy
signature where a photon is produced with no observable particle
balancing its transverse momentum.  Independently of the considered
new-physics model, the production of such heavy invisible particles
can be inferred from the missing transverse energy distribution, which
would exhibit an excess of events at high values as compared to the
SM.  Existing experimental searches for such an effect are, for
instance, presented in \citeres{LepmissingE,TevatronmissingE}.

Owing to their limited energy and luminosity, LEP2 and Tevatron only
provide weak constraints on anomalous couplings and production rates
for new particles
\cite{ewwg,TevatronZgWg,TevatronZgacop,TevatronWZacop}.  At the LHC,
the experimental collaborations will collect hundred thousands of
events coming from vector-boson pair production \cite{Haywood}. To
match the precision of the LHC experiments, the cross sections for
vector-boson pair-production processes have to be calculated beyond
leading order.  The next-to-leading-order (NLO) QCD corrections for
on-shell $\PW$ and $\PZ$ bosons were calculated in
\citeres{Smith:1989xz,Ohnemus:1992jn} and extended to include leptonic
decays in the narrow-width approximation and anomalous couplings in
\citeres{BaurWg,Ohnemus,BaurZg}. After the full NLO amplitudes
including leptonic decays had become available \cite{Dixon:1998py},
Monte Carlo programs incorporating these amplitudes have been
presented for $\PW\ga$ and $\PZ\ga$ production in \citere{DeFlorianWg}
and for $\PW\PW$, $\PW\PZ$, and $\PZ\PZ$ production in
\citeres{DixonWW,CampbellWW}. As a general feature, the QCD
corrections modify the leading-order di-boson cross sections at the
LHC by a positive amount of several tens of per cent, and have thus a
considerable impact on the measurement of gauge-boson couplings. At
this point, the question arises if the electroweak corrections to
vector-boson pair-production processes at the LHC lead to similarly
sizable effects. In the above-mentioned calculations typically only
universal electroweak corrections were taken into account such as the
running of $\alpha$ or corrections associated with the
$\rho$-parameter. This approach was based on the idea that
nonuniversal electroweak corrections only contribute significantly in
the high-energy domain, where the statistics at the LHC will be too
low to extract any physics from the experimental data.

However, this is in general not the case in the high-energy region,
which is of considerable experimental interest since as discussed
above the effects of the anomalous couplings and the effects of the
production of so far unknown heavy particles are most pronounced
there.  In fact, at high energies the electroweak corrections are
enhanced by so-called electroweak Sudakov logarithms, \ie double
logarithms of the process energy over the vector-boson mass
\cite{Beenakker:1993tt,Beccaria:1998qe,Ciafaloni:1998xg,Kuhn:1999de,Melles:2001ye}.
They can reach several tens of per cent and have to be taken into
account to make sure that an experimentally observed deviation from
the QCD-corrected SM predictions due to electroweak effects is not
misinterpreted as a signal for physics beyond the SM. The importance
of the electroweak corrections to gauge-boson pair-production
processes at the LHC has been confirmed by several calculations
\cite{Elena,Christoph,Kaiser}, which all point to a negative
contribution of some ten per cent, which well exceeds the statistical
errors.  Whereas the logarithmic electroweak corrections to $\PW\PW$,
$\PW\PZ$, $\PZ\PZ$ production in the high-energy limit have been
calculated including the corrections to the decay process as well as
the complete real corrections \cite{Kaiser}, for $\PW\gamma $
production only the logarithmic electroweak corrections to the
production subprocess are available \cite{Elena}.  The electroweak
corrections to $\PZ\gamma $ production have so far only been
determined for an on-shell $\PZ$~boson, \ie without taking into
account decay effects \cite{Christoph}.  Furthermore, the real
corrections for the latter two processes have been included only in
the soft and collinear limits or not at all. The goal of our work is
therefore the calculation of the virtual and real electroweak
corrections to the cross sections and distributions for the purely
leptonic channels $\Pp\Pp\to V\gamma+X\to l_1\bar{l}_2\ga+X$, where
$V=\PW,\PZ$.  Final states containing quarks are not discussed in this
work.

This paper is organized as follows: The strategy of our calculation is
described in \refse{sectionstrategy}. In \refse{se:input} we give 
the setup for the numerical evaluation.
Numerical results are presented in \refse{secresults}, and
\refse{secconclusion} contains our summary. Finally, some explicit
analytical results are listed in the appendices.

\section{Strategy of the calculation}
\label{sectionstrategy}

We consider the production of a photon along with a massive gauge
boson in proton--proton collisions, where the gauge boson decays
leptonically. The generic process reads
\beq
\Pp + \Pp \to V + \ga + X \to l_3 + \bar l_4 + \ga  + X,
\eeq 
where $\Pp$ denotes the incoming protons, $V$ indicates a $\PW$ or
$\PZ$~boson, $\ga$ the outgoing photon, $l_3$ the outgoing lepton,
$\bar l_4$ the outgoing anti-lepton, and $X$ the remnants of the
protons.

In the parton model, the corresponding cross section is obtained as a
convolution of the distributions $\Phi_{q_1|\Ppo}$ and
$\Phi_{\bar{q}_2|\Ppt}$ of the partons $q_1$ and $\bar{q}_2$ in the
incoming protons with the partonic cross sections
$\sigma_{q_1\bar{q}_2}$, averaged over spins and colours of the
partons,
\begin{equation}
\sigma_{{\rm \Ppo\Ppt}}(S)=\sum_{q_1,q_2}\int_0^1{\rm d}x_1
\int_0^1{\rm d}x_2\,  \Phi_{q_1|\Ppo}(x_1)
\Phi_{\bar{q}_2|\Ppt}(x_2)\sigma_{q_1\bar{q}_2}(\hat{s}). 
\label{hcs}
\end{equation}
The sum $\sum_{q_1,q_2}$ runs over all partonic initial states which
are allowed by charge conservation. In practice we include the quarks
and antiquarks $q_i =
\Pu,\bar\Pu,\Pd,\bar\Pd,\Ps,\bar\Ps,\Pc,\bar\Pc$. The integration
variables $x_i,i=1,2$ correspond to the partonic energy fractions.
Moreover, the quantities $S$ and $\hat{s}=x_1x_2S$ are the squared CM
energies of the hadronic process and the partonic subprocesses,
respectively.

For the partonic processes we adopt the generic notation
\begin{equation}
q_1(p_1,\si_1) + \bar{q}_2(p_2,\si_2)  \rightarrow V(p_3+p_4)
+\gamma(p_5,\la_5) \\ \rightarrow 
l_3(p_3,\si_3)+\bar{l}_4(p_4,\si_4)+\gamma(p_5,\la_5),
\label{qqVVffbara}
\end{equation}
where 
%$V=\PW,\PZ$, and $l_3,l_4=\nu_\Pe,\nu_{\mu},\nu_{\tau},\Pe,\mu$ and 
the arguments label the momenta and helicities $\si_i=\pm
1/2$, $i=1,\ldots,4$ and $\la_5=\pm1$  of the corresponding
particles. In the following, the masses of the external fermions are
denoted by $m_i$, $i=1,\ldots,4$. Since we treat all external fermions
as massless these appear only as regulators of mass singularities. The
charges of the fermions are denoted by $Q_i$, those of the
anti-fermions by $-Q_i$.

The lowest-order partonic cross sections are calculated
using the complete matrix elements for the process
\beq
q_1(p_1,\si_1) + \bar{q}_2(p_2,\si_2)  \rightarrow 
l_3(p_3,\si_3)+\bar{l}_4(p_4,\si_4)+\gamma(p_5,\la_5).
\label{qqffbara}
\eeq
This means that we include the full set of Feynman diagrams, in this
way accounting for all irreducible background coming from non-resonant
contributions. Analytic results are given in \refapp{secborn}.

The electroweak radiative corrections to \refeq{qqffbara}
consist of virtual corrections, resulting from loop diagrams, as well
as of real corrections, originating from the process
\begin{eqnarray}\label{eq:realprocess}
q_1(p_1,\sigma_1) + \bar q_2(p_2,\sigma_2) \to
 l_3(p_3,\sigma_3) + \bar l_4(p_4,\sigma_4) + \ga(p_5,\la_5) 
+ \gamma(p_6,\la_6)
\label{qqffbaraa}
\end{eqnarray}
with an additional photon with momentum $p_6$ and helicity
$\la_6=\pm1$.  Virtual and real corrections have to be combined
properly in order to ensure the cancellations of soft and collinear
singularities (\cf \refse{subsectionsoftcoll}). The helicity matrix
elements for the process \refeq{eq:realprocess} are listed in
\refapp{secbremsQED}.  For the calculation of the radiative
corrections we follow the approach used for the process
$\Pep\Pem\to\PWp\PWm\to4f$ in \citere{racoonWW}: The virtual
corrections are calculated in the {\em Leading-Pole Approximation}
(LPA), \ie we take only those terms into account that are enhanced by a
resonant massive gauge-boson propagator (\cf
\refse{subsectiondecays}), whereas the real corrections are calculated
from the full matrix elements for the processes
\refeq{eq:realprocess}.

\subsection{Absorption of initial-state mass singularities in parton
  distribution functions}
\label{subsectionhadcs}

At next-to-leading order in electroweak perturbation theory, the
partonic cross sections generally contain universal, initial-state
collinear singularities. The latter can be absorbed in the Parton
Distribution Functions (PDFs). To this end, one has to fix a
factorization scheme, and we choose the $\MSbar$ scheme, which defines
the connection between the lowest-order PDFs in (\ref{hcs}) and the
experimentally determined PDFs as \cite{Baur:1998kt,dittmaierkraemer}
\begin{multline}
\Phi_{q_i|\Pp}(x_i) =\overline{\Phi}^{\MSbar}_{q_i|\Pp}(x_i,Q^2)\\
-\frac{\alpha}{2\pi}Q_i^2 \int_{x_i}^1 \frac{{\rm d}z}{z}\;\overline{\Phi}^
{\MSbar}_{q_i|\Pp}\left(\frac{x_i}{z},Q^2\right)\left[\frac{1+z^2}{1-z}
\left(\ln\left(\frac{Q^2}{m_i^2}\right)-2\ln(1-z)-1\right)\right]_+.
\label{pdfMSbar}
\end{multline}
Here, we used the $(+)$-prescription
\begin{equation}
\int_a^1{\rm d}x \left[f(x)\right]_+g(x)=
\int_a^1{\rm d}xf(x)[g(x)-g(1)]
-g(1)\int_0^a{\rm d}xf(x), 
\label{plusprescription}
\end{equation}
and the quantity $Q$ denotes the electromagnetic factorization scale, which
should be set equal to the typical energy scale of the considered
process. Inserting the $\MSbar$ definition (\ref{pdfMSbar}) into the
hadronic cross section (\ref{hcs}) allows to write the latter
as a convolution of the experimentally defined PDFs with
the subtracted partonic cross section
\begin{equation}
\sigma_{{\rm \Ppo\Ppt}}(S,Q^2)=\sum_{q_1,q_2}\int_0^1{\rm
  d}x_1 \int_0^1{\rm d}x_2\,
\overline{\Phi}^{\MSbar}_{q_1|\Ppo}(x_1,Q^2)
\overline{\Phi}^{\MSbar}_{\bar{q}_2|\Ppt}(x_2,Q^2)
\sigma^{\IRsafe}_{q_1\bar{q}_2}(\hat{s},Q^2),
\label{sigmaPP}
\end{equation}
where 
\begin{equation}
\sigma^{\IRsafe}_{q_1\bar{q}_2}(\hat{s},Q^2)=
\sigma_{q_1\bar{q}_2}(\hat{s})-\sigma^{\MSbar}_{q_1\bar{q}_2,{\rm sing}}
(\hat{s},Q^2)+{\cal O}(\alpha^2)
\label{IRSafe}
\end{equation}
and
\begin{multline}
\sigma^{\MSbar}_{q_1\bar{q}_2,{\rm sing}}(\hat{s},Q^2)=\frac{\alpha}
{4\pi\hat{s}}\sum_{i=1,2} Q_i^2 \int_{0}^1{\rm d}z\int {\rm d}\Phi_i(z) 
\Abs{{\cal M}^{(0)}_{q_1\bar{q}_2}(\Phi_i(z))}\\
\times\frac{1}{z}\left[\frac{1+z^2}{1-z}\left(\ln\left(\frac{Q^2}{m_i^2}
\right)-2\ln(1-z)-1\right)\right]_+.
\label{sigmaMSbar}
\end{multline}
Here ${\cal M}^{(0)}_{q_1\bar{q}_2}$ denotes the lowest-order matrix
element, and $\Phi_i(z)$ represents the partonic phase-space
where the incoming particle $i$ has momentum $z p_i$.
After the subtraction, the resulting cross section is free of
collinear and soft singularities.
\par
A fully consistent inclusion of the electroweak corrections into the
calculation of the hadronic cross sections requires the use of PDFs in
which both photonic initial states and QED $\Oa$ corrections are taken
into account.  A first global analysis of parton distributions
incorporating QED contributions has been performed recently in
\citere{PDFQED}. It turns out that the photon PDFs are quite small
\cite{PDFQED}, and the QED corrections change the PDFs of the quarks
by less than 1\% \cite{Spiesberger,RothQEDPDF}.  These effects are
below the typical uncertainty of hadronic processes and below the
envisaged accuracy of our analysis. Accordingly, we restrict our
calculation to initial states containing quarks and use only
QCD-corrected PDFs in (\ref{sigmaPP}).  We nevertheless investigated
the dependence of our results on the electromagnetic factorization
scale $Q$ entering in \refeq{sigmaMSbar}, keeping
the QCD factorization scale Q that enters
$\overline{\Phi}^{\MSbar}_{q_i|\Ppo}(x_i,Q^2)$ in \refeq{sigmaPP}
fixed. For $\ga\PZ$ production, a variation of $Q$ by a factor 10
resulted in a change of the total cross section by less than 0.3\%, of
the $\ga\PZ$ invariant-mass distribution at large invariant mass by
less than 0.5\%, and of the transverse momentum distribution of the
photon at large transverse momenta by less than 1\%.

\subsection{Treatment of soft and collinear photon emission}
\label{subsectionsoftcoll}

In first-order electroweak perturbation theory, the subtracted partonic 
cross section receives contributions from the lowest order, the virtual 
corrections, the real corrections, as well as of the term (\ref{sigmaMSbar}) 
arising from the factorization of the initial-state collinear singularities:
\begin{equation}
\sigma_{q_1 \bar{q}_2}^{\IRsafe} (\hat{s},Q^2)= \sigma_{q_1 \bar{q}_2}^
{(0)}(\hat{s})+\sigma_{q_1 \bar{q}_2}^{{\rm virt}}(\hat{s})+
\sigma_{q_1 \bar{q}_2}^{{\rm real}}(\hat{s})-\sigma^{\MSbar}_{q_1\bar{q}_2,
{\rm sing}}(\hat{s},Q^2).
\label{partcs}
\end{equation}
In general, virtual and real corrections contain soft as well as
collinear singularities, which we regularize by introducing an
infinitesimal photon mass $\MA$ as well as small mass parameters $m_i$
for the external fermions.  These logarithmic singularities cancel
when adding real and virtual corrections, except for the collinear
logarithms arising from the initial state, which are absorbed in the
PDFs as described in \refse{subsectionhadcs}. Thus, all soft and
collinear singularities cancel in (\ref{partcs}), leaving the
subtracted cross section free of any mass regulators.
\par
However, this cancellation requires a careful treatment of soft and
collinear singularities.  In our work, we adopt the approach to divide
the virtual and real corrections into finite and singular parts, and
to cancel the dependence on the mass regulators by adding the singular
parts analytically. To extract the singular parts from the real
corrections, we employ two independent methods, the {\it
  phase-space-slicing} technique and the {\it dipole subtraction
  method}. In the following we give explicit expressions for our
treatment of infrared, \ie soft and collinear, singularities.

\subsubsection{Finite virtual corrections}

First, we define the finite virtual corrections by
\beqar\label{virtfin}
\sigma_{q_1\bar{q}_2,{\rm finite}}^{{\rm virt}}(\hat{s})&=& 
\frac{1}{2 \hat{s}}\int {\rm d}\Phi_0\, 
\Bigl(2\Re\left[{\cal M}_{q_1\bar{q}_2}^{(0)}(\Phi_0)
\left(\de{\cal M}_{q_1\bar{q}_2}^{{\rm
      virt}}(\Phi_0)\right)^{\star}\right]
\nl&&\qquad{}
-2\Re\left[{\cal M}_{q_1\bar{q}_2}^{(0)}(\Phi_0)
\left(\de{\cal M}_{q_1\bar{q}_2,\mathrm{sing}}^{{\rm
      virt}}(\Phi_0)\right)^{\star}\right]
\Bigr),
\label{virtsing}
\eeqar
where ${\cal M}_{q_1\bar{q}_2}^{(0)}$ and $\de {\cal
  M}_{q_1\bar{q}_2}^ {{\rm virt}}$ denote the matrix elements of the
Born process and the corresponding electroweak virtual corrections,
respectively, whereas $\Phi_0$ is the phase-space for the Born process
\refeq{qqffbara}. The subtracted singular parts are given by
\cite{racoonWW,Dittmaiersubtr}
\begin{equation}
2\Re\left[{\cal M}_{q_1\bar{q}_2}^{(0)}(\Phi_0)
\left(\de{\cal M}_{q_1\bar{q}_2,\mathrm{sing}}^{{\rm
      virt}}(\Phi_0)\right)^{\star}\right]
=\frac{\alpha}
{2\pi}\sum_{i,j=1\atop i\ne j}^n \tau_i\tau_jQ_iQ_j \Abs{{\cal M}_
{q_1\bar{q}_2}^{(0)}(\Phi_0)}({\cal L}(s_{ij},m_i^2)+C_{ij}),
\label{virtsingamp}
\end{equation}
where $n$ is the number of external particles of the lowest-order
process, $s_{ij}=(p_i+p_j)^2$, and $\tau_i=1$ for incoming particles and
outgoing anti-particles, whereas $\tau_i=-1$ for outgoing particles
and incoming anti-particles. The singularities are contained in 
\begin{equation}
{\cal L}(s_{ij},m_i^2)=\ln\left(\frac{\MA^2}{s_{ij}}\right)\left(1+\ln
\left(\frac{m_i^2}{s_{ij}}\right)\right)-\frac{1}{2}\ln^2\left(\frac{m_i^2}
{s_{ij}}\right)+\frac{1}{2}\ln\left(\frac{m_i^2}{s_{ij}}\right),
\label{Lsing}
\end{equation}
while the finite parts $C_{ij}$ are added for later convenience
and distinguish between incoming, $i,j=1,2$, and outgoing particles,
$i,j>2$,
\begin{equation}
C_{ij}=\left\{
\begin{array}{l l}
-\frac{\pi^2}{3}+\frac{3}{2},\qquad &i,j>2,\quad i\ne j,\\[1ex]
\phantom{-}\frac{\pi^2}{6}-1,&i=1,2,\quad j>2,\\[1ex]
-\frac{\pi^2}{2}+\frac{3}{2},&i>2,\quad j=1,2,\\[1ex]
-\frac{\pi^2}{3}+2,& i=1,2,\quad j=1,2,\quad i\ne j.\\[1ex]
\end{array}\right.
\label{Cij}
\end{equation}

\subsubsection{Phase-space slicing}

In short, the idea of the phase-space-slicing technique is to define
the finite real corrections by restricting the phase-space integration
to the region where the squared amplitude of the bremsstrahlung
process is finite. To this end, a technical cut $\dels$ on the
energy $E$ of the bremsstrahlung particle is introduced, as well as a
technical cut $\delc$ on the angles $\vartheta_i$ of the bremsstrahlung
particles with respect to all potential emitters.  The finite real
corrections are then given by
\begin{equation}
\sigma^{{\rm real,sli}}_{q_1\bar{q}_2,{\rm finite}}(\hat{s})=\frac{1}{2
\hat{s}}\int_{\parbox{1.9cm}{\tiny$E>\dels \sqrt{\hat{s}}/2$\\$
\cos\vartheta_i<1-\delc$}}
{\rm d}\Phi_{{\rm real}}\Abs{{\cal M}^{{\rm real}}_{q_1\bar{q}_2}
(\Phi_{{\rm real}})},
\label{realfinitePSS}
\end{equation}
where $\Phi_{{\rm real}}$ and ${\cal M}^{{\rm real}}$ denote the phase-space 
and the matrix element of the brems\-strahlung process \refeq{qqffbaraa}.

In the remaining singular parts of the real corrections, the momentum
of the bremsstrahlung particle is integrated out, up to a remaining
convolution over the momentum fraction $z$ of the initial state
particle after bremsstrahlung emission. Here the soft and collinear
singularities are again regularized by an infinitesimal photon mass
$\MA$ and by small mass parameters $m_i$ for the external fermions.
The resulting expressions \cite{racoonWW,Baur:1998kt} can be divided
into a part arising from initial-state radiation and a part arising
from final-state radiation. Adding them to the term (\ref{sigmaMSbar})
originating from the $\MSbar$ definition of the PDFs and to the
singular parts (\ref{virtsing}) of the virtual corrections, all
infrared
singularities cancel and one obtains a finite contribution
\begin{multline}
\label{sumsingPSS}
\sigma_{q_1\bar{q}_2,{\singf}}^{{\rm virt+real,
    sli}}(\hat{s},Q^2)=\sigma_{q_1\bar{q}_2,{\rm sing}}^{{\rm
    virt}}(\hat{s})+\sigma^{{\rm real,sli}}_{q_1\bar{q}_2,{\rm
    sing}}(\hat{s})-\sigma^{\rm \MSbar}_{q_1\bar{q}_2,{\rm
    sing}}(\hat{s},Q^2)\\ 
=\frac{1}{2\hat{s}}\int {\rm d}\Phi_0 
\left(\Abs{{\cal M}^{{\rm v+r, init}}_{q_1\bar{q}_2,{\singf}}(\Phi_0)}+\Abs{{\cal M}^{{\rm v+r, final}}_
{q_1\bar{q}_2,{\singf}}(\Phi_0)}\right)\\
+\frac{1}{2\hat{s}}\sum_{i=1}^2 \int \frac{{\rm d}z}{z}\int {\rm d}\Phi_i(z) \Abs{{\cal M}^{{\rm v+r,init,z}}_
{q_1\bar{q}_2,{\singf}}(\Phi_i(z))},
\end{multline}
where 
\begin{multline}
\Abs{{\cal M}^{{\rm v+r, init}}_{q_1\bar{q}_2,{\singf}}(\Phi_0)}=
\frac{\alpha}{2\pi}\sum_{i=1}^2 \sum_{{j=1 \atop j\ne i}}^n Q_i\tau_iQ_j\tau_j
\Abs{{\cal M}^{(0)}_{q_1\bar{q}_2}(\Phi_0)}\\\times\left[\frac{\pi^2}{3}-2+C_{ij}+{\rm Li}_2\left(1-\frac{4E_iE_j}
{s_{ij}}\right)\right.\\
\left.+\ln\left(\frac{\hat{s}}{s_{ij}}\frac{\delc}{2}\right)\left(\frac{3}{2}+\ln\dels^2\right)+\frac{1}{2}\ln^2
\left(\frac{\hat{s}}{s_{ij}}\right)\right],
\end{multline}
\begin{multline}
\Abs{{\cal M}^{{\rm v+r, final}}_{q_1\bar{q}_2,{\singf}}(\Phi_0)}=
\frac{\alpha}{2\pi}\sum_{i=3}^n \sum_{{j=1 \atop j\ne i}}^n Q_i\tau_iQ_j\tau_j
\Abs{{\cal M}^{(0)}_{q_1\bar{q}_2}(\Phi_0)}\\
\times \left[\pi^2-\frac{9}{2}+C_{ij}+{\rm Li}_2\left(1-\frac{4 E_i E_j}{s_{ij}}\right)\right.\\\left.
+\ln\left(\frac{4E_i^2}{s_{ij}}\frac{\delc}{2}\right)\left(\frac{3}{2}+\ln\left(\frac{\dels^2\hat{s}}{4E_i^2}
\right)\right)+\frac{1}{2}\ln^2\left(\frac{4E_i^2}{s_{ij}}\right)\right],
\end{multline}
\begin{multline}
\Abs{{\cal M}^{{\rm v+r,init,z}}_{q_1\bar{q}_2,{\singf}}(\Phi_i(z))}=-\frac{\alpha}{2\pi}\sum_{{j=1 \atop j\ne i}}^
n Q_i\tau_iQ_j\tau_j\Abs{{\cal M}^{(0)}_{q_1\bar{q}_2}(\Phi_i(z))}\\
\times \left[\frac{1+z^2}{1-z}\left(\ln\left(\frac{\hat{s}}{Q^2}\frac{\delc}{2}\right)+2\ln(1-z)+1\right)-\frac{2z}
{1-z}\right]_+ ,
\end{multline}
${\rm Li}_2$ denotes the dilogarithm
\begin{equation}
{\rm Li}_2(z)=-\int_0^z\;\frac{{\rm d}t}{t}\ln(1-t), \qquad |\arc(1-z)|<\pi,
\label{dilogdef}
\end{equation}
and $\Phi_i(z)$ represents the phase-space where the incoming particle
$i$ has momentum $z p_i$.

\subsubsection{Dipole subtraction method}

In the dipole subtraction method, the finite real corrections are
constructed by subtracting an auxiliary function from the squared
bremsstrahlung amplitude before integrating over phase-space,
\begin{equation}
\sigma^{{\rm real,sub}}_{q_1\bar{q}_2,{\rm finite}}(\hat{s})=\frac{1}
{2\hat{s}}\int{\rm d}\Phi_{{\rm real}}\left(\Abs{{\cal M}^{{\rm real}}_
{q_1\bar{q}_2}(\Phi_{{\rm real}})}-\Abs{{\cal M}^{\rm sub}_{q_1\bar{q}_2}
(\Phi_{{\rm real}})}\right).
\label{realfinitesubtr}
\end{equation}
The subtracted terms are added again after partial analytic integration
over the bremsstrahlung momentum.
The subtraction function has to be chosen such that it cancels all
soft and collinear singularities of the original integrand, so that
the difference \refeq{realfinitesubtr} can be integrated numerically
without regulators for these singularities. Moreover,
it has to be simple enough so that it can be integrated analytically
over the singular regions of phase space. We use the
process-independent {\em dipole subtraction formalism}, which was
introduced for massless QCD in \citeres{Catani1,Catani2,Catani3} and
generalized to massive fermions in \citere{Dittmaiersubtr}. We follow
the approach of \citere{Dittmaiersubtr} in the limit of small fermion
masses.

In the dipole subtraction formalism the subtraction function is
constructed from contributions that are labelled by ordered pairs $ij$
of charged fermions, so-called {\em dipoles}. The fermions $i$ and $j$
are called {\em emitter} and {\em spectator}, respectively. In the
formulation of \citere{Dittmaiersubtr} the subtraction function reads
\begin{equation}
\Abs{{\cal M}^{\rm sub}_{q_1\bar{q}_2}(\Phi_{{\rm real}})}= -4\pi \alpha 
\sum_{n_b=5}^{6}\sum_{i,j=1 \atop i\ne j}^n \tau_i\tau_j Q_i Q_j g_{ij}^
{\rm sub}(p_i,p_j,p_{n_b})\Abs{{\cal M}^{(0)}_{q_1\bar{q}_2}(\tilde{\Phi}^
{n_b}_{0,ij})}.
\label{Asubtr}
\end{equation} 
The sum over $n_b=5,6$ accounts for the fact that the 
photon appears twice
in the final state of the bremsstrahlung process (\ref{qqffbaraa}),
such that a separate subtraction term has to be introduced for each
final-state photon. Note that one of the two final-state photons has
always to be visible in the detector and thus does not give rise to
singularities. In what follows, we suppress the dependence of the
phase-space on $n_b$ and denote the bremsstrahlung momentum by
$k=p_{n_b}$. The phase-spaces $\tilde{\Phi}^{n_b}_{0,ij}$ in
(\ref{Asubtr}) are given by embedding prescriptions of the phase-space
$\Phi^{\rm real}$ of the bremsstrahlung process in the phase-space
$\Phi_0$ of the Born process. Both, the embedding prescription
$\tilde{\Phi}_{0,ij}^ {n_b}$ and the subtraction functions
$g_{ij}^{\rm sub}$ depend on the emitter $i$ and spectator $j$ being
incoming or outgoing. One therefore encounters four kinematically
different cases for the subtraction functions $g_{ij}^{\rm sub}$
\begin{equation}
g_{ij}^{\rm sub}(p_i,p_j,k)=\left\{
\begin{array}{l l}
\frac{1}{(p_i k)(1-y_{ij})}\left[\frac{2}{1-z_{ij}(1-y_{ij})}-1-z_{ij}\right],&\quad i,j>2,\\[1.5ex]
\frac{1}{(p_i k) x_{ij}}\left[\frac{2}{2-x_{ij}-z_{ij}}-1-z_{ij}\right],&\quad i>2,\quad j=1,2,\\[1.5ex]
\frac{1}{(p_i k) x_{ji}}\left[\frac{2}{2-x_{ji}-z_{ji}}-1-x_{ji}\right],&\quad i=1,2,\quad j>2,\\[1.5ex]
\frac{1}{(p_ik)v_{ij}}\left[\frac{2}{1-v_{ij}}-1-v_{ij}\right],&\quad
i=1,2,\quad j=1,2.
\end{array}\right.
\label{gsubtr}
\end{equation}
where
\begin{align}
x_{ij}&=\frac{p_ip_j+p_jk-p_ik}{p_ip_j+p_jk},& y_{ij}=\frac{p_ik}{p_ip_j+p_ik+p_jk},\nonumber\\
z_{ij}&=\frac{p_ip_j}{p_ip_j+p_j k},& v_{ij}=\frac{p_ip_j-p_ik-p_jk}{p_ip_j}.
\end{align}
Denoting the number of external particles by $n$, one has for the
embedding prescription if both the emitter and the spectator are part
of the final state
\begin{multline}
\tilde{p}^{\mu}_i=p_i^{\mu}+k^{\mu}-\frac{y_{ij}}{1-y_{ij}}p^{\mu}_j,\qquad 
\tilde{p}^{\mu}_j=\frac{1}{1-y_{ij}}p^{\mu}_j,\qquad  \tilde{p}^{\mu}_l=p^{\mu}_l,\\
i,j>2,\quad i\ne j, \quad l=1,\ldots,n,\quad l\ne i,j,
\label{momfinfin}
\end{multline}
whereas for an initial-state emitter and a final-state spectator one
gets
\begin{multline}
\tilde{p}^{\mu}_i=x_{ji} p^{\mu}_i,\qquad \tilde{p}^{\mu}_j=p^{\mu}_j+k^{\mu}-(1-x_{ji})p^{\mu}_i,\qquad  \tilde{p}^{\mu}_l=p^{\mu}_l,\\
i=1,2,\quad j>2, \quad l=1,\ldots,n,\quad l\ne i,j.
\label{mominfin}
\end{multline}
The embedding prescription for a final-state emitter and an
initial-state spectator can be obtained from (\ref{mominfin}) by
exchanging $i\leftrightarrow j$ everywhere. Finally, in case of an
initial-state emitter and an initial-state spectator the embedding
prescription is given by
\begin{multline}
\tilde{p}_i^{\mu}=v_{ij}p_i,\qquad \tilde{p}_j=p_j,\qquad
\tilde{p}_l^{\mu}=\Lambda_{\;\;\nu}^{\mu}p_l^{\nu},\qquad
i,j=1,2,\quad i \ne j, \quad l=3,\ldots,n
\label{mominin}
\end{multline}
with the boost matrix
\begin{equation}
\Lambda_{\;\;\nu}^{\mu}=g_{\;\;\nu}^{\mu}-\frac{(P_{ij}+\tilde{P}_{ij})^{\mu}(P_{ij}+\tilde{P}_{ij})_{\nu}}{P_{ij}^2+P_{ij}\tilde{P}_{ij}}+\frac{2\tilde{P}_{ij}^{\mu}P_{ij,\nu}}{P_{ij}^2}.
\end{equation}
Here, $\tilde{P}_{ij}$ represents the total initial-state momentum
after the projection,
$\tilde{P}_{ij}^{\mu}=\tilde{p}_i^{\mu}+\tilde{p}_j^{\mu}$, and
\begin{equation}
P_{ij}=p_1+p_2-k=\sum_{k=3}^n p_k.
\end{equation}

The subtracted contribution can be integrated over the (singular)
photonic degrees of freedom up to a remaining convolution over a
variable $z$. In this integration the regulators $\MA$ and $m_i$ must
be retained, and soft and collinear singularities appear as logarithms
in these mass regulators. The resulting expressions
\cite{racoonWW,Dittmaiersubtr} can again be divided into a part
arising from initial-state radiation and a part arising from
final-state radiation. Adding them to the term (\ref{sigmaMSbar})
originating from the $\MSbar$ definition of the PDFs and to the singular
parts (\ref{virtsing}) of the virtual corrections, the infrared
singularities cancel and one finds
\begin{multline}
\sigma_{q_1\bar{q}_2,{\singf}}^{{\rm virt+real,sub}}(\hat{s},Q^2)=\sigma_{q_1\bar{q}_2,{\rm sing}}^{{\rm virt}}(\hat{s})
+\sigma^{{\rm real,sub}}_{q_1\bar{q}_2,{\rm sing}}(\hat{s})-\sigma^{\rm \MSbar}_{q_1\bar{q}_2,{\rm sing}}(\hat{s},Q^2)\\
=\frac{1}{2\hat{s}}\sum_{i=1}^2 \int \frac{{\rm d}z}{z}\int {\rm d}\Phi_i(z) \left(\Abs{{\cal M}^{{\rm v+r,init,z}}_
{q_1\bar{q}_2,{\singf}}(\Phi_i(z))}+\Abs{{\cal M}^{{\rm v+r,
    final,z}}_{q_1\bar{q}_2,{\singf}}(\Phi_i(z))} 
\right),
\label{sumsingsubtr}
\end{multline}
where 
\begin{multline}
\Abs{{\cal M}^{{\rm v+r,init,z}}_{q_1\bar{q}_2,{\singf}}(\Phi_i(z))}=-\frac{\alpha}{2\pi} \sum_{{j=1 \atop j\ne i}}
^n Q_i\tau_iQ_j\tau_j\Abs{{\cal M}^{(0)}_{q_1\bar{q}_2}(\Phi_{i}(z))}\\\times \left[\frac{1+z^2}{1-z}\left(\ln\left
(\frac{\vert\tilde{s}_{ij}\vert}{Q^2z}\right)+2\ln(1-z)\right)+1-z\right.\\
\left.-\delta_{j>2}\left(
\frac{2}{1-z}\ln(2-z)-(1+z)\ln(1-z)\right)
\right]_+,
\label{ampsquinsubtr}
\end{multline}
\begin{multline}
\Abs{{\cal M}^{{\rm v+r,final,z}}_{q_1\bar{q}_2,{\singf}}(\Phi_i(z))}\\
=\frac{\alpha}{2\pi} Q_i\tau_i(\tau_1Q_1+\tau_2 Q_2) \Abs{{\cal M}^{(0)}_{q_1\bar{q}_2}(\Phi_i(z))}\left[\frac{1}{1-z}\left(2\ln
\left(\frac{2-z}{1-z}\right)-\frac{3}{2}\right)\right]_+,
\label{ampsqufinalsubtr}
\end{multline}
and $\delta_{j>2}=1$ for $j>2$ and $\delta_{j>2}=0$ for $j\le2$.  Note
that the finite part \refeq{Cij} of the singular virtual corrections
\refeq{virtsingamp} has been chosen to exactly cancel the end-point
parts resulting from the subtraction function (\ref{Asubtr}) such that
only $z$-dependent contributions remain in \refeq{ampsquinsubtr} and
\refeq{ampsqufinalsubtr}.

\subsubsection{Master formula}

Altogether, in our approach the subtracted partonic cross section,
which is free of soft and collinear singularities,
can be written as a sum of four different parts
\begin{equation}
\sigma_{q_1\bar{q}_2}^{\IRsafe}(\hat{s},Q^2)= \sigma_{q_1\bar{q}_2}^{(0)}(\hat{s})+\sigma_{q_1\bar{q}_2,{\rm finite}}
^{{\rm virt}}(\hat{s})+\sigma_{q_1\bar{q}_2,{\rm finite}}^{{\rm real}}(\hat{s})+\sigma_{q_1\bar{q}_2,{\rm sing}}^
{{\rm virt+real}}(\hat{s},Q^2),
\label{Masterformula}
\end{equation}
where the finite virtual corrections in the second term of the right-hand 
side are defined by (\ref{virtsing}). The last two terms are given by
(\ref{realfinitePSS}) and (\ref{sumsingPSS}), respectively, if using
the phase-space-slicing technique and by (\ref{realfinitesubtr}) and
(\ref{sumsingsubtr}), respectively, in case of using the subtraction
method. 

\subsection{Treatment of finite-width effects}
\label{subsectiondecays}

The processes (\ref{qqVVffbara}) involve the production and decay of
an unstable particle $V=\PW,\PZ$. The corresponding propagator leads
to a pole in the amplitude, which has to be regularized by
incorporating the finite width of the massive vector bosons. Several
methods to include finite-width effects in a perturbative calculation
have been discussed in the literature
\cite{Stuart,Aeppli,Aeppli2,Beenakkerfermloop,BeenakkerPole,ComplexMS,PassarinoWidth,ElenaWidth,BeenakkerWidth1,BeenakkerWidth2,BenekePole,BenekeWidth}.
The simplest approach is given by the fixed-width scheme, which
corresponds to replacing the resonant propagator by one containing a
constant finite width.  Since the finite width results from
resummation of an incomplete set of higher-order contributions, its
introduction potentially violates gauge invariance. For Born
amplitudes, the effects of this violation have been shown to be
numerically small in the fixed-width scheme (see \eg
\citeres{Beenakkerfermloop,ComplexMS,Roth,Argyres}) by comparing the
results to those obtained with other manifestly gauge-invariant
prescriptions, as \eg the complex-mass scheme \cite{ComplexMS}.
\par
In our calculation, following the approach of \citere{racoonWW}, we
adopt the fixed-width scheme to evaluate the Born cross section and
the real corrections.  For the virtual corrections, on the other hand,
we use the LPA \cite{racoonWW,Aeppli,BeenakkerPole}. In this
approximation only the leading term in an expansion about the
resonance pole, \ie the residue divided by the resonant propagator, is
kept. Since the residue, which is related to physical amplitudes for
on-shell production and decay, and the resonant propagator are
gauge-invariant, the LPA yields a gauge-invariant result. As all
non-resonant contributions are neglected, the number of contributing
diagrams is considerably reduced, and only loop integrals with up to
four propagators appear.  The error induced by using the LPA for the
virtual corrections can be estimated to be of the order
$\alpha\Ga_V/M_V$, where $M_V$ and $\Ga_V$ are the mass and the width
of the vector boson $V$, respectively, and thus of the order of a few
per mille. This estimate was confirmed for single W production at
hadron colliders in \citere{dittmaierkraemer} and for WW production in
electron positron annihilation in
\citeres{Denner:2005es,Denner:2005fg}. Thus, the LPA should be more
than sufficient for the considered process, as long as the resonant
diagrams dominate.

\par 
Following the approach presented in \citere{racoonWW}, the virtual
corrections evaluated in LPA can be further split into
{\it factorizable} and {\it non-factorizable} parts, such that
the finite virtual corrections can be written as
\begin{multline}
\sigma_{q_1\bar{q}_2,{\rm finite}}^{{\rm virt},{\rm LPA}}(\hat{s})
= \frac{1}{2 \hat{s}}\int {\rm d}\Phi_0\,
2\Re\left[{\cal M}^{(0),{\rm LPA}}_{q_1\bar{q}_2}(\Phi_0,\Phi_0^{\rm osh})\right.
\\
\times\left.\left(\de{\cal M}^{{\rm virt},{\rm LPA}}_{q_1\bar{q}_2,{\rm
    fac}}(\Phi_0,\Phi_0^{\rm osh})
+\de{\cal M}_{q_1\bar{q}_2,{\rm nfac}}^
{{\rm virt},{\rm LPA}}(\Phi_0,\Phi_0^{\rm osh})
-\de{\cal M}_{q_1\bar{q}_2,{\rm sing}}^
{{\rm virt},{\rm LPA}}(\Phi_0,\Phi_0^{\rm osh})
\right)^\star
\right].
\label{virtfiniteLPA}
\end{multline}
In \refeq{virtfiniteLPA}, 
${\cal M}^{(0),{\rm LPA}}_{q_1\bar{q}_2}(\Phi_0,\Phi_0^{\rm osh})$ denotes
the Born matrix element in LPA
\begin{equation}
{\cal M}^{(0),{\rm LPA}}_{q_1\bar{q}_2}(\Phi_0,\Phi_0^{\rm osh})=\frac{R^{(0)}(\Phi_0^{\rm osh})}{(p_3+
p_4)^2-M_V^2+\ri M_V\Gamma_V},
\label{BornLPA}
\end{equation}
which receives contributions from all Born diagrams containing the
resonant propagator. The residue $R$ of the amplitude at the pole is
evaluated using a phase-space $\Phi^{\rm osh}_0=\{p_i^{\rm
  osh},i=1,\ldots,5\}$ projected on the mass-shell of the decaying
particle, \ie $(p_3^{\rm osh}+p_4^{\rm osh})^2=M_V^2$. We specify our
choice for the on-shell projection in \refapp{app-projection}.
The factorizable virtual corrections in \refeq{virtfiniteLPA}
get contributions from all diagrams where the resonant propagator
appears outside the loop, \ie from diagrams that factorize into a
production part, a decay part, and the resonant propagator. In LPA,
they can be written as
\begin{equation}
\de{\cal M}_{q_1\bar{q}_2,{\rm fac}}^{{\rm virt},{\rm
    LPA}}(\Phi_0,\Phi_0^{\rm osh})=\frac{R_{{\rm fac}}^{{\rm virt}} 
(\Phi_0^{\rm osh},\MA)}{(p_3+p_4)^2-M_V^2+\ri M_V\Gamma_V},
%,\qquad  (p_3^{\rm osh}+p_4^{\rm osh})^2=M_V^2,
\label{evalfac}
\end{equation}
where all soft singularities are regularized by the infinitesimal photon mass 
parameter $\MA$.
\par 
The non-factorizable virtual corrections receive contributions from 
all diagrams where a massless virtual particle, in our case a photon, 
connects the production and decay subprocesses or one of these
subprocesses to the resonance. The infrared singularities originating
from photon emission off the unstable boson $V$ are here regularized
by keeping the resonant momentum off-shell wherever the on-shell limit
leads to a singularity. The contribution of the diagrams where the
photon couples to the resonance are already partly contained in the
factorizable virtual corrections as defined above, such that this
contribution has to be subtracted from the non-factorizable
corrections to avoid double counting \cite{racoonWW}.  As an
advantage, the non-factorizable virtual corrections can be evaluated
in the {\it Extended Soft Photon Approximation} (ESPA), which
corresponds to setting the loop momentum to zero wherever this does
not lead to a resonance.  The non-factorizable corrections
\begin{multline}
\de{\cal M}_{q_1\bar{q}_2,{\rm nfac}}^{{\rm virt},{\rm LPA}}(\Phi_0,\Phi_0^{\rm osh})=\frac{R_{{\rm nfac}}^{{\rm virt},
{\rm ESPA}}(\Phi_0^{\rm osh},(p_3+p_4)^2-M_V^2)}{(p_3+p_4)^2-M_V^2+\ri M_V\Gamma_V}\\
=\frac{1}{2}{\cal M}^{(0),{\rm LPA}}_{q_1\bar{q}_2}(\Phi_0,\Phi_0^{\rm osh}) \delta^{{\rm virt}}_{{\rm nfac}}
(\Phi_0^{\rm osh},(p_3+p_4)^2-M_V^2),
\end{multline}
are then proportional to the Born matrix element in LPA. We give the 
analytical expression for the correction factor 
$\delta^{{\rm virt}}_{{\rm nfac}}$ in \refapp{secnfac}.
\par
Note that in order not to spoil the cancellation of soft and collinear
singularities between the virtual corrections in LPA and the real corrections 
evaluated exactly, we consistently subtract in (\ref{virtfiniteLPA}) the 
singular virtual corrections evaluated in LPA
\begin{multline}
2\Re \left[{\cal M}^{(0),\mathrm{LPA}}_{q_1\bar{q}_2}(\Phi_0,\Phi_0^{\rm osh})
 \left(\de {\cal M}_{q_1\bar{q}_2,{\rm sing}}^{{\rm virt, LPA}}(\Phi_0,\Phi_0^{\rm osh})\right)^\star\right]\\
=\frac{\alpha}{2\pi}\sum_{i,j=1\atop 
i\ne j}^n \tau_i\tau_jQ_iQ_j \Abs{{\cal M}_{q_1\bar{q}_2}^{(0),{\rm LPA}}(\Phi_0,\Phi_0^{\rm osh})}({\cal L}
(s^{\rm osh}_{ij},m_i^2)+C_{ij}),
\label{virtsingLPA}
\end{multline}
but we use the exact singular virtual corrections in (\ref{sumsingPSS}) and 
(\ref{sumsingsubtr}). In (\ref{virtsingLPA}),
$s_{ij}^{\rm osh}=(p_i^{\rm osh}+p_j^{\rm osh})^2$, and ${\cal L}$ and
$C_{ij}$ are given by (\ref{Lsing}) and (\ref{Cij}), respectively.

\subsection{Implementation}

We have implemented our strategy in a Mathematica package called {\tt
  Pole}, which works as an extension of the computer-algebra packages
{\tt FeynArts3} \cite{FeynArts} and {\tt FormCalc3.1} \cite{FormCalc}. To
be explicit, {\tt Pole} extends these packages by the following
features:
\begin{itemize}
\begin{item}
amplitude generation for initial-state hadrons,
\end{item}
\begin{item}
  amplitude generation for a given intermediate state to select only
  resonant factorizable and resonant non-factorizable diagrams,
\end{item}
\begin{item}
evaluation of the non-factorizable part of the amplitude in ESPA
\end{item}
\begin{item}
reduction of fermion chains using the Weyl--van der Waerden formalism,
\end{item}
\begin{item}
numerical evaluation of hadronic cross sections,
\end{item}
\begin{item}
numerical evaluation of the virtual corrections in LPA,
\end{item}
\begin{item}
  phase-space integration for an arbitrary number of external
  particles using the generic phase-space generator of {\tt Lusifer}
  \cite{Lusifer},
\end{item}
\begin{item}
numerical evaluation of the real corrections using phase-space slicing
or the dipole subtraction method.
\end{item}
\end{itemize}
The main changes for the amplitude generation are the possibility to
specify hadronic initial states as well as the resonant particles in
the intermediate state. The latter information is used to only
generate the resonant factorizable and the resonant non-factorizable
diagrams. In the \FC\ part, we have implemented the Weyl--van der
Waerden formalism, which reduces the fermionic spinor structures down
to Weyl-spinor products and Kronecker deltas in the helicities. Note
that our implementation of the Weyl--van der Waerden method works
independently of the formalism contained in {\tt FormCalc4}
\cite{FormCalc4}, where the fermionic structures are numerically
evaluated as two-dimensional matrix--vector products. As an advantage
of our approach with regard to {\tt FormCalc4}, the matrix element is
split into helicity amplitudes, and only non-vanishing components are
calculated.

As a consequence, the kinematical dependence of the amplitudes as
calculated by {\tt Pole} is completely different from the amplitude
given back by {\tt FormCalc3.1}, so that we had to rewrite the 
{\tt Fortran} part of {\tt FormCalc3.1}.  The input required by the
new code are model parameters, kinematical cuts, PDF set, as well as
the aforementioned on-shell projection of the outgoing momenta.
Using this input, the program automatically calculates hadronic
cross sections and kinematical distributions including real QED
corrections as well as virtual electroweak corrections evaluated in
LPA. The phase-space integration is done using the generic Monte Carlo
generator of {\tt Lusifer} 
\cite{Lusifer}, which is able to handle the kinematics of a process
with an arbitrary number of final-state particles. To flatten the
propagator peaks, {\tt Lusifer} employs the multi-channel importance
sampling
\cite{Berends:1984gf,Berends:1986ig,Hilgart:1992xu,Berends:1994pv}
with adaptive weight optimization
\cite{aprioriweights}.
Since the generator was originally designed to calculate Born
cross sections only, we had to extend it to incorporate an appropriate
treatment of soft and collinear singularities. We implemented both,
the dipole subtraction method and the phase-space-slicing technique as
presented in \refse{subsectionsoftcoll}.
\par 
{\tt Pole} is designed to handle processes involving an arbitrary
final state and an arbitrary number of resonances. The
applicability is limited to processes where the decay products are
stable, \ie it cannot handle cascade decays where unstable particles
decay into unstable particles.  Furthermore, the treatment of the
non-factorizable corrections as well as the implementation of the
phase-space slicing and subtraction methods are limited to massless
external particles.

\section{Input and checks}
\label{se:input}
\subsection{Parameter input and process definition}
\label{secinput}

We consider three classes of processes:
\begin{alignat}{5} \label{PPnnbara}
\Pp + \Pp  &\rightarrow&
\nu_l+\bar{\nu}_l&{}+\gamma + X ,&\qquad &l=\Pe,\mu,\tau,&\\
\label{PPllbara}
\Pp + \Pp  &\rightarrow&
\,l\;+\,\bar{l}\;&{}+\gamma + X,&\qquad &l=\Pe,\mu,&\\
\Pp + \Pp  &\rightarrow& \nu_l
+\,\bar{l}\;&{}+\gamma + X,&\qquad &l=\Pe,\mu.&
\label{PPnlbara}
\end{alignat}
The first two classes allow to analyse $\PZ\gamma$ production, while
the third one contains $\PW\gamma$ as intermediate state.  In the
following sections, we present results for the LHC at CM energy
$\sqrt{S}=14\TeV$ and an integrated luminosity $L=100\fb^{-1}$ per
experiment.  We neglect all fermion masses except for the mass of the
top quark. We though keep the fermion masses in arguments of loop
integrals as a regularization parameter for possible collinear
divergences, where we use the values \cite{PDBook,CDFmt}
\begin{align}
\Mu&=0.066\GeV,& \Mc&= 1.6\GeV,&\Mt&=178\GeV,\nonumber\\
\Md&=0.066\GeV,& \Ms&=0.15\GeV,&\Mb&=4.9\GeV,\nonumber\\
\Me&=5.109989\times10^{-4}\GeV,& m_{\mu}&=0.105658369\GeV,&m_{\tau}&=1.77699\GeV.
\label{fermionmasses}
\end{align} 
For the vector-boson masses and decay widths we take \cite{PDBook}
\begin{eqnarray}
\MW=80.425\GeV, \qquad \GW=2.124\GeV,\nl
\MZ=91.1876\GeV, \qquad \GZ=2.4952\GeV,
\end{eqnarray}
and the Higgs-boson mass is fixed to  
\begin{equation}
\MH=115\GeV.
\end{equation} 

We use the $\GF$ scheme, \ie we define the fine structure constant in terms 
of the Fermi constant 
\begin{equation}
\alpha=\alpha_{\GF}=\frac{\sqrt{2}}{\pi} \GF \MW^2 \sw^2,
\label{alphaGF}
\end{equation}
where the quantity $\sw^2=1-\MW^2/\MZ^2$ denotes the sine of the weak
mixing angle squared, and we use the value $\GF= 1.16637\times
10^{-5}\;{\rm GeV^{-2}}$ for the Fermi constant.  The definition
(\ref{alphaGF}) effectively resums contributions associated with the
evolution of $\alpha$ to the $\PW$-boson mass and incorporates leading
universal $\Mt$-dependent two-loop corrections. For the coupling of
one real final-state photon, we use instead the value
$\alpha=\al(0)=1/137.035999$, \ie we rescale the cross section with 
$\al(0)/\al_{\GF}$.
Thus we have for the
lowest-order cross section $\si^{(0)}$ and for the corresponding
corrections $\si^{(1)}$:
\beq
\si^{(0)}\propto \al(0)\alpha_{\GF}^2, \qquad \si^{(1)}\propto \al(0)\alpha_{\GF}^3.
\eeq
The terms absorbed in
$\alpha_{\GF}$ have to be subtracted from the finite virtual
corrections (\ref{virtfiniteLPA}) calculated in the $\al(0)$ scheme,
once for each vertex parametrized by $\alpha_{\GF}$, to avoid double
counting. Since in our case we have two such
couplings in the lowest-order cross section, the finite virtual corrections are evaluated according to
\begin{equation}   
\sigma_{q_1\bar{q}_2,{\rm finite},\alpha_{\GF}}^{\virt,{\rm LPA}}(\hat{s})=\sigma_{q_1\bar{q}_2,{\rm finite}}^{\virt,
{\rm LPA}}(\hat{s})-2\sigma_{q_1\bar{q}_2}^{(0),{\rm
  LPA}}(\hat{s})\Delta r^{(1)}. 
\label{subtrdR}
\end{equation}
The one-loop contribution $\Delta r^{(1)}$ to $\De r$ can for instance
be found in \citere{Denner:1991kt}.
%\begin{multline}
%\Delta r^{(1)}=\left.\frac{\partial \Sigma^{\gamma}(k^2)}{\partial k^2}\right|_{k^2=0}-\frac{\cw^2}{\sw^2}\left
%(\frac{\Sigma^{Z}(\MZ^2)}{\MZ^2}-\frac{\Sigma^{W}(\MW^2)}{\MW^2}\right)+\frac{\Sigma^{W}(0)-\Sigma^W(\MW^2)}{\MW^2}
%\vspace*{0.4cm}\\
%+2\frac{\cw}{\sw}\frac{\Sigma^{\gamma Z}(0)}{\MZ^2}+\frac{\alpha}{4\pi \sw^2}\left(6+\frac{7-4 \sw^2}{2 \sw^2}\ln \cw^2
%\right).\nonumber
%\end{multline}
%Here $\Sigma^V, V=\gamma,Z,W$ are the vector-boson self-energies and
%$\Sigma^{\gamma Z}$ denotes the photon--$\PZ$-boson mixing-energy.
\par
We neglect all loop corrections to the quark
mixing, and simply multiply the cross sections of the partonic
subprocesses with the squares of the quark-mixing matrix elements
\begin{align}
V_{\Pu\Pd}&=0.974,&V_{\Pu\Ps}&=\sqrt{1-V_{\Pu\Pd}^2},&V_{\Pu\Pb}&=0,\nonumber\\
V_{\Pc\Pd}&=-\sqrt{1-V_{\Pu\Pd}^2},&V_{\Pc\Ps}&=0.974,&V_{\Pc\Pb}&=0,\nonumber\\
V_{\Pt\Pd}&=0,&V_{\Pt\Pc}&=0,&V_{\Pt\Pb}&=1
\end{align}
for $\PW\gamma$-production processes
\par
Following our discussion at the end of Section \ref{subsectionhadcs}, we 
neglect the QED corrections to the PDFs. We therefore choose the only 
QCD-corrected CTEQ6M set \cite{Pumplin} to calculate the hadronic 
cross sections and we neglect all contributions of the initial states 
containing bottom quarks. Denoting the momenta of the incoming protons
by $P_{1,2}$ and the final-state momenta collectively by $p_{\rm
  fin}$, the hadronic cross section for the $\PZ\gamma$-production
processes \refeq{PPnnbara} and \refeq{PPllbara} can be written as
\begin{multline}
\sigma_{\Pp\Pp}(S,Q^2)=\int_0^1{\rm d}x_1 \int_0^1{\rm d}x_2\;
\frac{1}{2x_1x_2S}\int{\rm d}\Phi(p_{\rm fin}) \\
\times \sum_{q=\Pu,\Pd,\Pc,\Ps} \left(\Abs{{\cal M}_{q\bar{q}}(x_1P_1,x_2P_2,p_{
\rm fin})}\overline{\Phi}^{\MSbar}_{q|\Pp}(x_1,Q^2)
\overline{\Phi}^{\MSbar}_{\bar{q}|\Pp}(x_2,Q^2)
\right.\\\left.
{}+{}\Abs{{\cal M}_{q\bar{q}}(x_2P_2,x_1P_1,p_{\rm fin})}\overline{\Phi}^{\MSbar}_{q
|\Pp}(x_2,Q^2)\overline{\Phi}^{\MSbar}_
{\bar{q}|\Pp}(x_1,Q^2)\right),
\end{multline}
and we use ${\cal M}_{\Pc\bar{\Pc}}={\cal M}_{\Pu\bar{\Pu}}$ and ${\cal
  M}_{\Ps\bar{\Ps}}={\cal M}_{\Pd\bar{\Pd}}$ in the zero mass limit
for the external fermions. For the $\PW\gamma$-production process
\refeq{PPnlbara}, the effects of the quark mixing are taken into
account by setting the quark-mixing matrix to the unit matrix for the
calculation of the amplitudes, and by convoluting the squared
amplitudes according to
\begin{multline}
\sigma_{\Pp\Pp}(S,Q^2)=\int_0^1{\rm d}x_1 \int_0^1{\rm
  d}x_2\;\frac{1}{2x_1x_2S} 
\int{\rm d}\Phi(p_{\rm fin}) \\\times
\sum_{q=\Pu,\Pc}\sum_{q^{\prime}=\Pd,\Ps} \Abs{V_{qq^{\prime}}}\left(\Abs{{\cal M}_{q\bar{q}^{\prime}}(x_1P_1,x_2P_2,p_{\rm fin})}
\overline{\Phi}^{\MSbar}_{q|\Pp}(x_1,Q^2)\overline{\Phi}^{\MSbar}_{\bar{q}'|\Pp}(x_2,Q^2)\right.\\\left.
{}+{}\Abs{{\cal M}_{q\bar{q}'}(x_2P_2,x_1P_1,p_{\rm fin})}\overline{\Phi}^{\MSbar}_{q|\Pp}(x_2,Q^2)\overline{\Phi}^{\MSbar}_
{\bar{q}'|\Pp}(x_1,Q^2)\right),
\end{multline}
and we use ${\cal M}_{\Pu\bar{\Pd}}={\cal M}_{\Pu\bar{\Ps}}={\cal
  M}_{\Pc\bar{\Pd}}={\cal M}_{\Pc\bar{\Ps}}$ for unit quark-mixing matrix. 
\par
Finally, for the typical energy scale $Q$ used to evaluate the PDFs we
choose \cite{Haywood}
\begin{equation}
Q^2=\frac{1}{2}\left(M_V^2+p_{{\rm T},V}^2+p_{{\rm
      T},\gamma}^2\right),\qquad V=\PZ,\PW, 
\label{muF}
\end{equation}
where $p_{{\rm T},V}$ denotes the transverse momentum of the resonance
\begin{equation}
p_{{\rm T},V}=\left\{
\begin{array}{c l}
p_{\rm T}^{{\rm miss}} &{\rm for\quad }\Pp \Pp  \rightarrow \nu_l \bar{\nu}_l \gamma,\\[1ex]
\sqrt{(p_{{\rm T},x}^{l}+p_{{\rm T},x}^{\bar{l}})^2+(p_{{\rm
      T},y}^{l}+p_{{\rm T},y}^{\bar{l}})^2}\qquad &{\rm for\quad }
\Pp \Pp \rightarrow l \bar{l} \gamma,\\[1ex]
\sqrt{(p_{{\rm T},x}^{l}+p_{{\rm T},x}^{{\rm miss}})^2+(p_{{\rm
      T},y}^{l}+p_{{\rm T},y}^{{\rm miss}})^2}\qquad &{\rm for\quad }
\Pp \Pp  \rightarrow \nu_l \bar{l}\gamma.\\
 \end{array}
\right.
\label{PTV}
\end{equation}
The missing transverse momentum $p_{\rm T}^{{\rm miss}}$ is here
defined as the sum of the transverse momenta of all particles visible
in the detector, whereas $p_{{\rm T},x}^{{\rm miss}}$ and $p_{{\rm
    T},y}^{{\rm miss}}$ denote missing momentum in $x$ and $y$
direction, respectively. We use the same scale $Q$ in the calculation
of the electromagnetic $\MSbar$ subtraction term \refeq{sigmaMSbar}.
\par 

We have implemented a general set of cuts, proper for LHC analyses.
In order to specify this set, we introduce the azimuthal--pseudorapidity
separation between particle $i$ and $j$
\begin{equation}
\Delta R_{i j}=\sqrt{(\eta_i-\eta_j)^2+(\varphi_i-\varphi_j)^2},
\label{Rrec}
\end{equation}
where  $\varphi_i$ is the azimuthal angle parametrizing the
spatial part of the external momentum $p_i$, and
the corresponding pseudo-rapidity 
\begin{equation}
\eta_i=-\ln\left(\tan\frac{\theta_i}{2}\right)
\end{equation}
is a measure of the production angle $\theta_i$ of the considered
particle with respect to the beam.

In order to define observables that are free of soft and collinear
singularities, we use the following recombination scheme:
\begin{itemize}
\item Photons close to the beam, \ie with a rapidity
  $\vert\eta^{\gamma}\vert>\eta^{{\rm rec}}=2.5$, or with too small
  energy, \ie $E^{\gamma}<E^{\rm{rec}}=2\GeV$, are treated as
  invisible, \ie they only contribute to the missing momentum.
\item Photons with $\vert\eta^{\gamma}\vert<\eta^{{\rm rec}}=2.5$ and
  $E^{\gamma}>E^{\rm{rec}}=2\GeV$ which are close to a final state
  charged lepton or another photon, \ie which fulfil $\Delta R_{\gamma
    i}<R^{{\rm rec}}=0.1$ for at least one $i=e,\mu,\gamma$, are
  recombined.  This means that the momentum of the photon and the
  momentum $p_i$ of the final-state charged lepton or
  photon for which $\Delta R_{\gamma i}$ has the lowest value are
  added and considered as an effective lepton or photon momentum.
\item Photons with $\vert\eta^{\gamma}\vert<\eta^{{\rm rec}}=2.5$,
  $E^{\gamma}>E^{\rm{rec}}=2\GeV$, and $\Delta R_{\gamma i}>R^{{\rm
      rec}}=0.1$ for all $i=e,\mu,\gamma$ are treated as visible.
\end{itemize}

\par 
For the acceptance cuts, we consider an event to contribute to the
cross section if the momenta for the leptons after possible photon
recombination and the momentum of at least one of the visible
final-state photons fulfil the following requirements:
\begin{itemize}
\begin{item}
  $p_{\rm T}^{\gamma}>p_{\rm T}^{\gamma,{\rm c}}=50\GeV\;(100\GeV)$
  for at least one photon in case of\/ $\PW\gamma$ ($\PZ\gamma$)
  production,
\end{item}
\begin{item}
$p_{\rm T}^l>p_{\rm T}^{l,{\rm c}}=20\GeV$ for all final-state charged leptons,
\end{item}
\begin{item}
  $\Delta R_{ij}>R^{\rm c}=0.7$, $i,j=l,\gamma$, for the
  rapidity--azimuthal angle separation (\ref{Rrec}) between two charged
  leptons and between charged leptons and visible photons.
\end{item}
\begin{item}
  $\vert\eta\vert<\eta^{\rm c}=2.5$ for at least one of the
  final-state photons and all final-state charged leptons.
\end{item}
\begin{item}
  $p_{\rm T}^{{\rm miss}}>p_{\rm T}^{{\rm miss},{\rm
      c}}=50\GeV\;(100\GeV)$ for the missing transverse momentum in
  case of $\PW\gamma$ ($\PZ\gamma$) production for final states
  containing at least one neutrino,
\end{item}
\end{itemize} 
Furthermore, we introduce cuts on the invariant masses of the resonant
gauge bosons. These cuts enhance the contributions of the
gauge-boson pair-production process with respect to background
diagrams and simultaneously reduce the error of the LPA.  We impose
the so-called reconstruction cut
\begin{equation}
\MZ-20\GeV <\sqrt{(p_3+p_4)^2}< \MZ+20\GeV
\end{equation}
on the phase-space for the process (\ref{PPllbara}).  For the case of
$\PW\gamma $-production \refeq{PPnlbara}, the neutrino momentum is not
an observable, such that one has to rely on the missing
transverse momentum to restrict the squared energy of the lepton pair
to the phase-space region around the resonance. To this end, a cut on
the transverse mass
\begin{equation}
M_{\rm T}^{l\nu}< \MW+20\GeV
\end{equation}
is appropriate, where the transverse mass for a lepton--neutrino pair
$l\nu_l$ is defined as 
\begin{equation}
M_{\rm T}^{l\nu}=\sqrt{(\vert p^{\rm miss}_{{\rm T}}\vert+\vert p_{{\rm T},l}\vert)^2-(p^{\rm miss}_{x}+p_{x,l})^2-
(p^{\rm miss}_{y}+p_{y,l})^2}.
\label{MTrdef}
\end{equation}
For the case of the $\PZ$ boson decaying into two neutrinos
\refeq{PPnnbara} none of the decay products is visible, and thus
neither of the above reconstruction cuts can be imposed.

For the bremsstrahlung process the scale $Q$ is fixed using the
possibly recombined momenta as follows.  If both photons are visible,
we use
\begin{equation}
Q^2=\frac{1}{2}\left(M_V^2+p_{{\rm T},V}^2+p_{{\rm T},\gamma_1}^2+p_{{\rm T},\gamma_2}^2\right),\qquad V=\PZ,\PW,
\label{muFbrems}
\end{equation} 
to evaluate the PDFs. If only one photon is visible, the expression
used for the scale $Q$ depends on the considered final state. For the
$l\bar{l} \gamma\gamma$ final state, we reconstruct the transverse
momentum of the bremsstrahlung particle from momentum conservation and
use the expression (\ref{muFbrems}). When the final state contains one
or more neutrinos, the bremsstrahlung momentum contributes to the
missing momentum and we use (\ref{muF}) and (\ref{PTV}) instead.

\subsection{Checks of the calculation}
\label{sectests}
% based on 10^8 events for virtual and 10^8 for real

To check the amplitudes generated by \Pole\ numerically, we performed
an explicit paper-and-pencil calculation of the Born amplitudes, the
non-factorizable virtual corrections, and the real corrections. We
have listed the corresponding expressions in \refapp{explicitamps}.
In addition, we compared the Born amplitude and the bremsstrahlung
amplitude for zero width numerically to the result of Madgraph
\cite{Madgraph,mgraph2} for some phase-space points and found complete
agreement.  For a check of the amplitude for the factorizable virtual
corrections, we generated the diagrams for the partonic processes
corresponding to \refeq{qqffbara} by means of \FA. We then switched
off the non-resonant and non-factorizable diagrams by hand, and
translated the corresponding amplitudes into {\tt Fortran} code using
{\tt FormCalc4}. In the so-generated code, we replaced the resonant
propagator by the corresponding propagator including a finite width.
The resulting amplitudes were checked by comparing to the output of
\Pole\ for some set of phase-space points, as well as by comparing the
integrated cross sections, yielding in each case complete agreement.
\par
The numerical evaluation of the finite virtual corrections
\refeq{virtfiniteLPA} was tested by numerically varying the UV
regulator, \ie the mass parameter $\mu$ for the dimensional
regularization, the infrared regulator, \ie the infinitesimal photon
mass $\MA$, and the masses of the external fermions.  The finite
virtual corrections were found to be completely independent of these
parameters, and we fixed the UV regulator to $\mu=1\GeV$, the infrared
regulator to $\MA=1\GeV$, and the masses of the external fermions to
the values listed in (\ref{fermionmasses}).
\begin{figure}
\epsfig{file=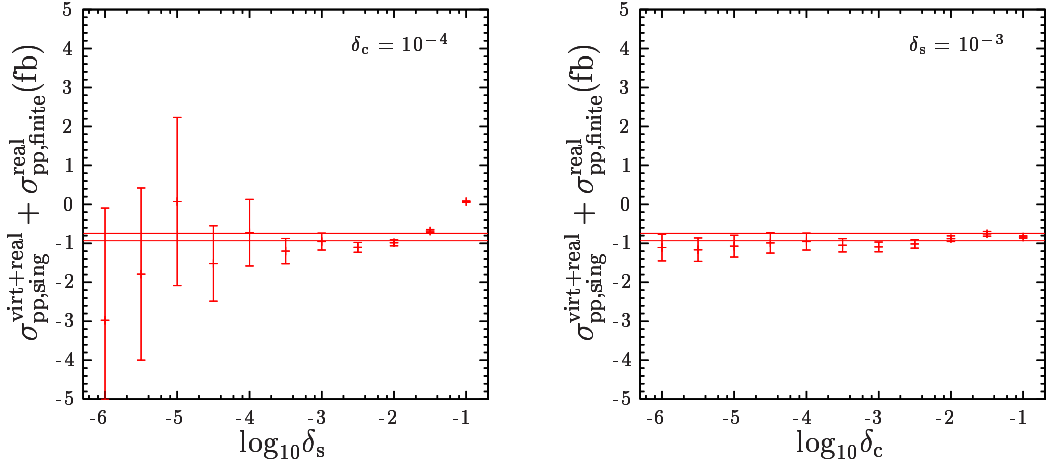} 
\caption{\label{deltacstest} Dependence
  of the sum of the singular virtual corrections, the complete real
  corrections, and the term arising from the $\MSbar$ definition of
  the PDFs for $\Pp\Pp \rightarrow \nu_l\bar{l}\gamma + X$ on the
  phase-space-slicing cut parameters. The horizontal lines indicate
  the error bars of our results obtained by using the subtraction
  method.}
\end{figure}
\par
To check the treatment of the real corrections, we calculated the
dependence of the sum of (\ref{realfinitePSS}) and (\ref{sumsingPSS})
on the slicing cut parameters $\dels$ and $\delc$ and compared the
corresponding values to the sum of (\ref{realfinitesubtr}) and
(\ref{sumsingsubtr}) obtained from the dipole subtraction method.
In \reffi{deltacstest} we show the dependence on  $\dels$ for fixed
$\delc=10^{-4}$  and the dependence on  $\delc$ for fixed
$\dels=10^{-3}$ for the process $\Pp\Pp \rightarrow \nu_l\bar{l}\gamma
+ X$.  The error bars reflect the uncertainty of the Monte Carlo
integration.  The cross section shows a plateau over the full
considered range for $\delc$, whereas the variation with $\dels$ leads
to a plateau in the region $10^{-5}<\dels<10^{-2}$. In addition, the
numerical values obtained in these ranges agree with those obtained by
using the subtraction method within the integration error, which we
indicate in \reffi{deltacstest} by the horizontal lines. The same
analysis for the other considered processes yielded similar results.
For the results to be presented in the next section, we accordingly
fixed the slicing cuts to $\dels=10^{-3}$ and $\delc=10^{-4}$.
\par
Finally, tree-level amplitudes and phase-space integration have been tested 
by comparing the outcome of our 
code with the results obtained with 
the program used for the analyses in \citere{Elena}. This latter code employs
matrix elements generated by means of {\tt PHACT}\cite{phact}, a set of 
routines based on the helicity-amplitude formalism of \citere{habm}, and an 
independent integration method. For the sake of comparison, the original 
version of the program was extended to include the $\PZ\ga$-production 
processes (\ref{PPnnbara}) and (\ref{PPllbara}). The results for all three 
Born cross sections agree at the per-mille level.  Furthermore, we
numerically cross-checked the lowest-order results for the
distributions to be presented in the next section for all three
processes. Complete agreement has been reached, which confirms
our implementation of phase-space integration, cut routines,
and histogram generation.

\section{Numerical results}
\label{secresults}
% based on 2x10^9 events

In this section, we illustrate the effect of the electroweak
corrections on hadronic $\PW\gamma$ and $\PZ\gamma$
production at the LHC. For the process \refeq{PPnnbara} we sum over all
three neutrino flavours, while for the two processes \refeq{PPllbara}
and \refeq{PPnlbara} we include the first two lepton families only.

In the definition of the distributions for these processes,
the possible presence of two photons in the final state, one coming
from the real corrections, can lead to ambiguities.  When observables
of the final-state photon are involved, our prescription is to
attribute the event weight to the bin corresponding to the photon with
the largest energy. 

\subsection{Total cross sections}

\begin{table}[t]
\begin{center}
\begin{tabular}{|l| |c| |c| c|| c|}
%\hline
%\multicolumn{5}{|c|}{{\bf The Results for the Total Hadronic Cross-Sections}}\\
\hline
\hline
&$\sigma_{\Pp\Pp}^{(0)}$(fb)&\multicolumn{2}{|c||}{$\sigma^{(1),{\rm tot}}_{\Pp\Pp}$(fb)}&$1/\sqrt{2 L \sigma^{(0)}_
{\Pp\Pp}}\;(\%)$\\
\hline
\hline
$\nu_l\bar{\nu}_l\gamma$&{\footnotesize $212.26(7)$}&{\footnotesize $-9.65(3)$}&{\footnotesize $-4.5\%$}&
{\footnotesize $0.5\%$}\\
\hline
$l\bar{l}\gamma$&{\footnotesize $~38.99(9)$}&{\footnotesize $-2.61(7)$}&{\footnotesize $-6.7\%$}&{\footnotesize $1.1\%$}\\
\hline
$\nu_l\bar{l}\gamma$&{\footnotesize  $124.57(8)$}&{\footnotesize $-2.36(5)$}&{\footnotesize $-1.9\%$}&{\footnotesize 
$0.6\%$}\\
\hline
\end{tabular}
\caption{Total lowest-order cross section (second column) as well as
  the electroweak $\Oa$ corrections in absolute size (third column)
  and in per cent of the lowest-order cross section (fourth column)
  for the three considered final states.  The last column shows the
  statistical error for an integrated luminosity of $L=2\times
  100/{\rm fb}$.}
\label{sigmatot}
\end{center}
\end{table}
In \refta{sigmatot}, we show the total hadronic cross section for the
three processes (\ref{PPnnbara})--(\ref{PPnlbara}).  The second column
contains the lowest-order results, while the third and fourth entries
display the value of the $\Oa$ corrections and their contribution
relative to the lowest-order cross section, respectively.  For the
three considered processes, the electroweak corrections are negative
and of the order of $-2\%$ to $-7\%$.  This is to be compared with the
last column of \refta{sigmatot}, where we show an estimate of the
statistical error based on an integrated luminosity $L=100\,{\rm
  fb}^{-1}$ for two experiments. As can be seen, the size of the
electroweak corrections well exceeds the statistical uncertainty.
They are of the same order of magnitude as the systematic error, which
is expected to be dominated by the error on the PDFs and thus
presently of the order of 5\% for processes initiated by quarks. The
systematic uncertainty could possibly be reduced by considering ratios
of cross sections.  On the other hand, the size of the $\Oa$
corrections is strongly cut dependent and in fact increases with
growing $\CM$ energy and vector-boson scattering angle.  Larger values
for the electroweak corrections are found by choosing, for instance, a
more stringent cut on the transverse momentum of the photon. Such a
constraint is useful for new-physics searches, although decreasing the
statistics (see \citere{Elena} and references therein).  Also,
distributions can be differently affected by the radiative corrections
according to the specific observable at hand. In the following
sections we illustrate this point for some sample variables.

\subsection{Distributions for $\Pp\Pp\rightarrow \nu_l\bar{\nu}_l\gamma + X$}
\begin{figure}[t]
\epsfig{file=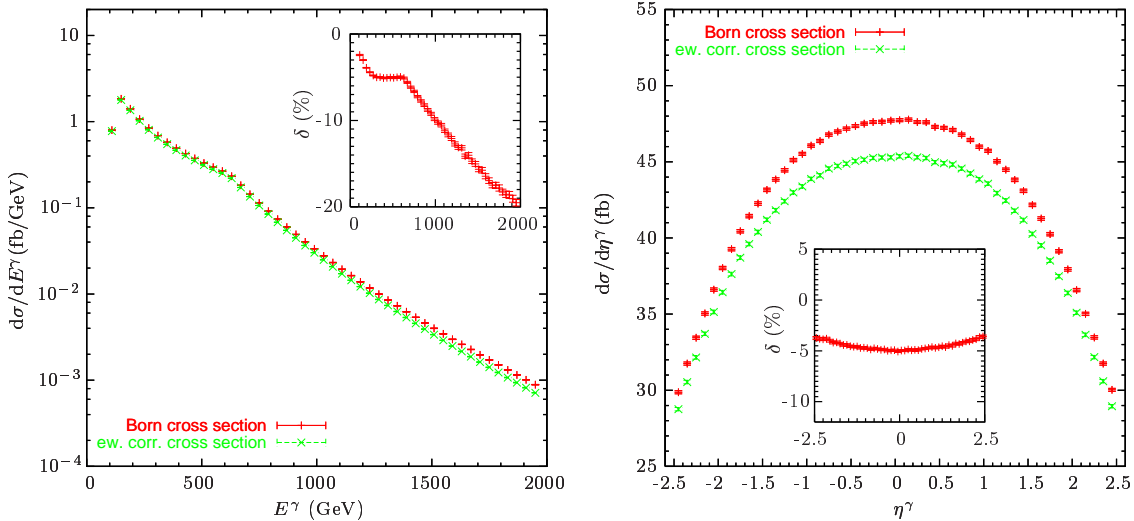, scale = 0.92}
\caption{\label{resultsnunu} Distributions in energy (left plot) and rapidity (right plot) of the final-state photon 
  for the process $\Pp\Pp\to \nu_l\bar{\nu_l}\gamma+X$,
  $l=\Pe,\mu,\tau$.  The inserted plots show the relative $\Oa$
  corrections in per cent normalized to the lowest order.}
\end{figure}
We begin by considering distributions for $\Pp\Pp\rightarrow
\nu_l\bar{\nu}_l\gamma + X$. The error bars in the following plots
result from the Monte Carlo integration errors.
In the left plot of \reffi{resultsnunu}, we show the energy
distribution of the final-state photon.  As previously mentioned, this
distribution can be used to probe the existence of new particles which
escape detection. These particles, once produced along with the
photon, would lead to an excess of events in the high-energy domain.
The kink at $E^{\gamma}=600\GeV$ is due to the cut
$\vert\eta\vert<\eta^{\rm c}$ on the rapidity of the final-state
photon.  This cut restricts the cosine of the photon production angle
$\theta^{\gamma}$ in the laboratory frame to the interval
\begin{equation}
\frac{\exp^{-2\eta^{\mathrm{c}}}-1}{\exp^{-2\eta^{\mathrm{c}}}+1}<\cos\theta^{\gamma}<\frac{\exp^{2\eta^{\mathrm{c}}}-1}
{\exp^{2\eta^{\mathrm{c}}}+1},
\end{equation}
such that for our choice $\eta^{\mathrm{c}}=2.5$:
\begin{equation}
\frac{p_{\rm T}^{\gamma}}{E^{\gamma}} = \vert \sin\theta^{\gamma}\vert> 0.163071\ldots\approx\frac{1}{6}.
\end{equation}
Since we also applied a cut of $p_{\rm T}^{\ga,\mathrm{c}}=100\GeV$ on the
transverse momentum of the photon, this condition is always fulfilled
for $E^{\gamma}\lsim600\GeV$ and the rapidity cut is never applied for
these photon energies. For $E^{\gamma}\gsim600\GeV$, on the other hand,
the rapidity cut will start to exclude more and more events, which
leads to a steeper decrease of the Born and NLO distributions.
Assuming an LHC luminosity of $L=2\times 100/{\rm fb}$, the bin
$E^{\gamma}=800\pm 10\GeV$ collects a contribution of ${\rm
  d}\sigma/{\rm d} E^{\gamma}\approx 5\times 10^{-2}\,{\rm fb}/{\rm
  GeV}$ to the total cross section, and will thus contain about 
\begin{equation}
N_{\rm ev}\approx L\times \frac{{\rm d}\sigma}{{\rm
    d}E^{\gamma}}\Delta E^{\gamma}=200\,{\rm fb}^{-1}\times
\frac{5\times 10^{-2}\,{\rm fb}}{{\rm GeV}}\,{20\GeV}=200 
\label{nev}
\end{equation}
events, where $\Delta E^{\gamma}=20\;{\rm GeV}$ is the bin width. In
this energy region, the electroweak corrections are of the order of
$-10\%$. The increase of the corrections with $E^{\gamma}$ 
can be attributed to large logarithms of Sudakov type. The electroweak
corrections should therefore be included to match the experimental
accuracy. This is the first calculation of the $\Oa$ corrections to
the photon spectrum for $\nu_l\bar{\nu_l}\gamma$ final states. Up to
now, only QCD-corrected distributions have been computed.  Since the
electroweak corrections reduce cross sections and distributions,
considering only lowest-order or QCD-corrected results could
overestimate the SM background when looking for new particles.  An
excess of events in the high-energy domain could then be
misinterpreted as compatible with the SM predictions, and could
therefore be missed.
\par
The right plot of \reffi{resultsnunu} shows the distribution in the
photon rapidity.  The electroweak corrections are here of the order of
$-3\%$ to $-5\%$, and are thus smaller as in the case discussed above
and of the order of the present systematic uncertainty from PDFs.
This is because all bins receive the dominant contributions from
events with low CM energies, where the electroweak corrections are
small. An estimate of the event rate as in \refeq{nev} applied to the
bin $\eta^{\gamma}=0\pm 0.05$ gives 900 events.

\subsection{Distributions for $\Pp\Pp\rightarrow l\bar{l}\gamma + X$}

\begin{figure}[htbp]
\begin{center}
\epsfig{file=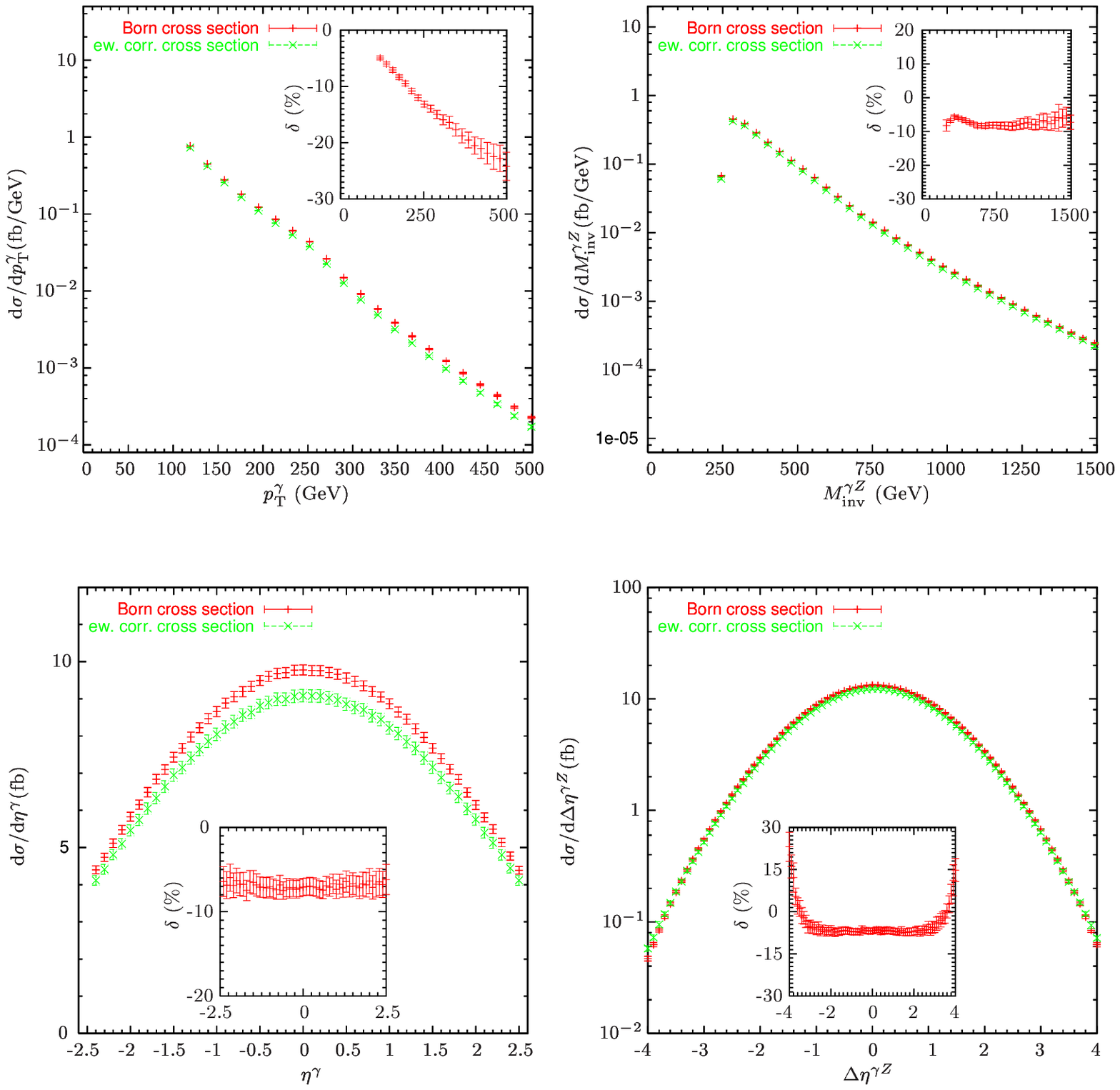, scale=0.92}
\end{center}
\caption{\label{resultsee1} Distributions in the transverse momentum
  of the final-state photon (upper left plot), in the $\gamma\PZ$
  invariant mass (upper right plot), in the photon
  rapidity (lower left plot), and in the photon--$\PZ$-boson rapidity
  difference (lower right plot) for $\Pp\Pp\rightarrow l\bar{l}\gamma + X$, 
  $l=\Pe,\mu$.  The
  inserted plots show the relative $\Oa$ corrections in per cent
  normalized to the lowest order.}
\end{figure}
We turn now to the process (\ref{PPllbara}), mediated by $\PZ\gamma$
production, and we show in \reffi{resultsee1} some sample
distributions.  The upper left plot displays the photon transverse
momentum distribution.  The kink at $p_{\rm T}^{\gamma}=250\GeV$ can
be explained by the cut $\De R_{l\bar l} > R^{\mathrm{c}}$ on the
rapidity-azimuthal angle separation of the two leptons resulting from
the \PZ~boson.  The position of the kink is approximately given by
%\beq
% p_{\rm T}^{\gamma} = p_{\rm T}^{ R^{\mathrm{c}}} \equiv
% \MZ\sqrt{\frac{1+\cos R^{\mathrm{c}}}{1-\cos 
%  R^{\mathrm{c}}}}.
%\eeq
\beq
 p_{\rm T}^{\gamma} = p_{\rm T}^{ R^{\mathrm{c}}} \equiv
\frac{2\MZ}{\sqrt{\vphantom{A^A}2 \exp(R^{\mathrm{c}})-1}-1}.
\eeq
For $p_{\rm T}^{\PZ}=p_{\rm T}^{\gamma}<p_{\rm T}^{ R^{\mathrm{c}}}$
the $ R^{\mathrm{c}}$ cut is not effective, while for $p_{\rm
  T}^{\PZ}=p_{\rm T}^{\gamma}>p_{\rm T}^{ R^{\mathrm{c}}}$ the decay
products of the $\PZ$ boson are so strongly boosted that this cut
eliminates many of the events.  The effects of the electroweak
corrections to this distribution have already been studied in
\citere{Christoph} for an on-shell final-state $\PZ$ boson, and with
the real corrections included in the soft and collinear limits only.
In this case, the corrections were found to be large and negative, of
the order of $-20\%$ and to increase in absolute size with rising
$p_{\rm T}^{\gamma}$. As can be seen from \reffi{resultsee1}, our
calculation largely confirms these results.  Since the effects of the
anomalous couplings are most pronounced in the high-$p_{\rm T}$
region, the electroweak corrections could modify the experimental
sensitivity to possible new physics. For the process considered here,
the QCD corrections to the distribution in the photon transverse
momentum have been calculated in \citeres{Ohnemus,BaurZg,DeFlorianWg}.
If applying a jet veto, they are found to range between $+60\%$ and
$-20\%$ \cite{DeFlorianWg}. At NLO, the contributions arising from
electroweak and QCD corrections are thus comparable in magnitude.
Including the electroweak radiative effects in the experimental
analysis is therefore mandatory in order not to overestimate the SM
background.
\par
In the upper right plot of \reffi{resultsee1}, we show the
distribution in the final-state invariant mass.  A simple estimate as
in \refeq{nev} shows that the process should be observable up to
$M_{\rm inv}^{\gamma\PZ}=750$--$1000\GeV$. As can be seen from the
inset plot, the total electroweak corrections amount to roughly
$-8\%$, independently of the $M_{\rm inv}^{\gamma\PZ}$ value.
Electroweak radiative-correction effects are thus relevant for this
important distribution.  Earlier results for the electroweak
corrections to the invariant-mass distribution have been published in
\citere{Christoph} under the aforementioned approximations. Large
negative corrections of the order of $-15\%$ were found, increasing in
magnitude towards larger invariant masses.  Comparing these results to
the curve shown in the inset of the upper right plot in
\reffi{resultsee1}, we rather find the relative size of the virtual
and real corrections approaching a plateau with rising invariant mass.
We attribute this effect to angular-dependent logarithms, which are
enhanced for events with photons collinear to the beams and compensate
the Sudakov logarithms. Such events are eliminated by a cut on $p_{\rm
  T}^{\gamma}$ and thus do not influence the $p_{\rm T}^{\gamma}$
distribution, but contribute to the invariant-mass distribution also
for large invariant masses.  Thus, compared to the results published
in \citere{Christoph}, the inclusion of the real corrections reduces
the impact of the electroweak corrections for large invariant masses.

The distribution in the photon rapidity is shown in the lower left
plot of \reffi{resultsee1}. The relative size of the electroweak
corrections can be seen from the inserted plot to be of the order of
$-7\%$.  Finally, the photon--$\PZ$-boson rapidity difference is shown
in the lower right plot of \reffi{resultsee1}. While the corresponding
corrections are at a similar level as for the two previous
distributions in the central region, they become large and positive
near $\vert \Delta\eta^{\gamma\PZ}\vert\approx4$.  For $\vert
\Delta\eta^{\gamma\PZ}\vert>5$, the cut
$\vert\eta^{\gamma}\vert<\eta^{\mathrm{c}}=2.5$ eliminates all events
with only one photon and a pair of leptons in the final state, and
only bremsstrahlung events contribute. Thus, the smallness of the
lowest order cross section causes the large relative corrections near
and above $\vert \Delta\eta^{\gamma\PZ}\vert\approx4$.  However, the
event rate expected at the LHC in this region is too small for
experimental observation. The remaining central part of the curve does
not present novelties compared to the two previous distributions.

\subsection{Distributions for $\Pp\Pp\rightarrow {\nu}_l\bar{l}\gamma + X$}
\begin{figure}[htbp]
\begin{center}
\epsfig{file=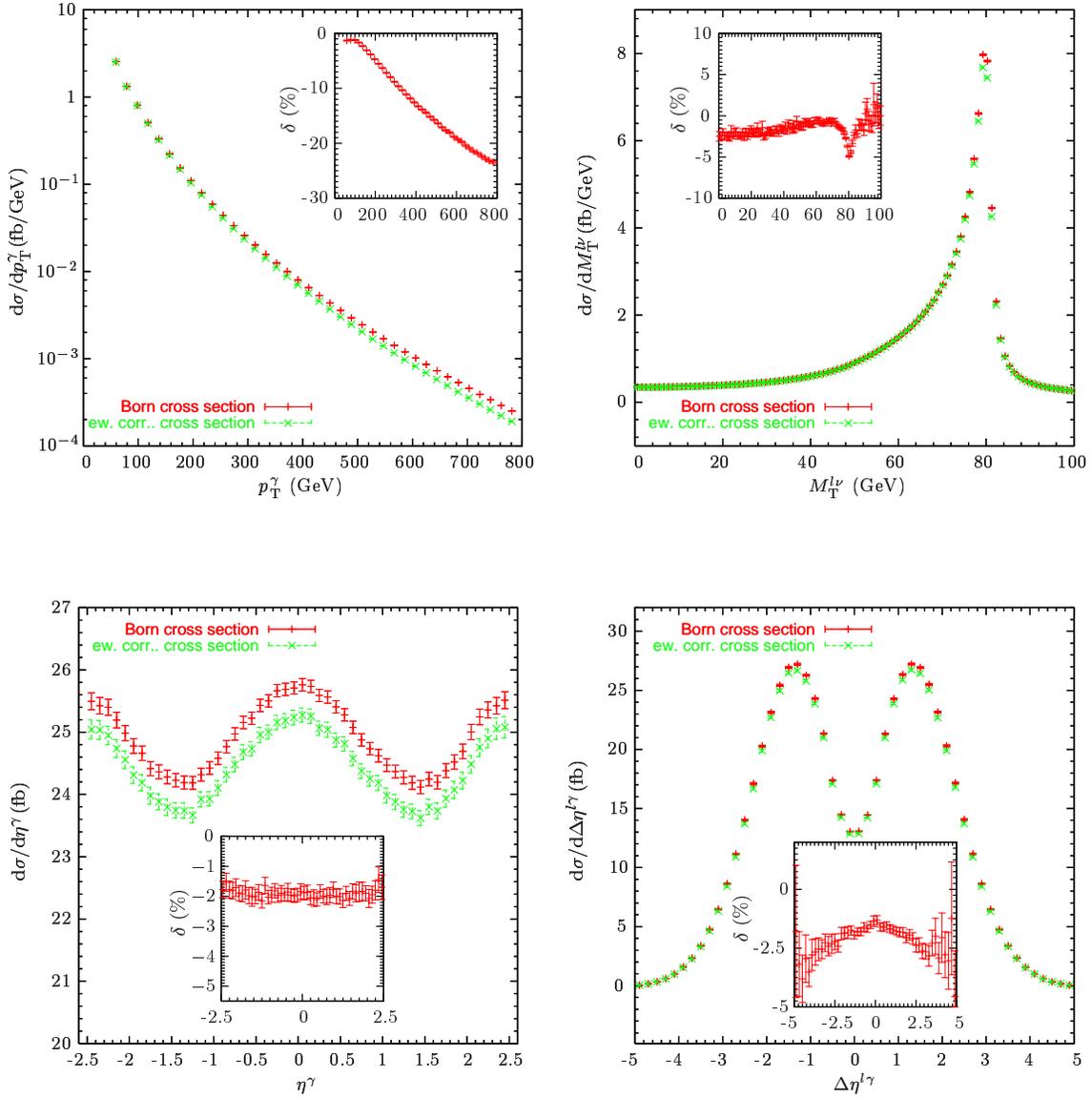, scale=0.92}
\end{center}
\caption{\label{resultsenu1} Distributions in the transverse momentum
  of the final-state photon (upper left plot), in the transverse mass
  of the lepton--neutrino pair (upper right plot), in the photon
  rapidity (lower left plot) and in the lepton--photon rapidity
  difference (lower right plot) for 
  $\Pp\Pp\rightarrow \nu_l \bar{l} \gamma + X$, $l=\Pe,\mu$. The
  inserted plots show the relative electroweak corrections in per cent
  normalized to the lowest order.}
\end{figure}
Turning next to the $\PW\gamma$ production process, we show in
\reffi{resultsenu1} some distributions for $\Pp\Pp\rightarrow
{\nu}_l\bar{l}\gamma + X$, $l=\Pe,\mu$.  Starting with the
transverse-momentum distribution of the photon in the upper left plot,
an estimate as in \refeq{nev} shows that the event rate should be
observable up to $p_{\rm T}^{\gamma}=300$--$400\GeV$. An analysis of
the effects of the electroweak corrections on this distribution summed
over $\nu_l\bar{l}$ and $l\bar{\nu}_l$ final states has already been
published in \citere{Elena}, taking into account the $\PW$-boson decay
but including only the leading logarithmic virtual corrections to the
production subprocess in the high-energy approximation.  Large
negative corrections were found, which can reach up to $-12\%$ at
$p_{\rm T}^{\gamma}=400\GeV$. As can be seen from the inset plot, we
reproduce this result.  At $p_{\rm T}^{\gamma}=400\GeV$, the
electroweak corrections are found to contribute about $-13\%$ of the
Born result, which is well above the systematic and statistical
errors. The QCD corrections to the transverse-momentum distribution of
the photon have been analysed in \citere{DeFlorianWg}.  If applying a
jet veto, large positive corrections of up to $60\%$ for low
transverse momenta have been found. They decrease to $-5\%$ to $-10\%$
of the Born cross section at $p_{\rm T}^{\gamma}=400\GeV$. For high
transverse momenta, the electroweak corrections are thus of the same
order of magnitude or even larger than the QCD corrections, and
further reduce the lowest-order distribution. Again, a data analysis
performed just including QCD corrections could overestimate the SM
background in anomalous coupling measurements.

In the upper right plot of \reffi{resultsenu1}, we display our
results for the distribution in the transverse mass of the final-state
lepton--neutrino pair as defined in (\ref{MTrdef}). The curves
terminate at the cut value $M_{\rm T,c}^{l\nu}=\MW+20\;{\rm
  GeV}\approx 100.4\GeV$. The $\PW$-boson resonance can be observed as
a pronounced peak, an effect which could be used to determine the
$\PW$-boson mass as a consistency check. 
 The electroweak corrections can be seen from the inserted plot
to contribute most at the peak, where they amount to $-4\%$
of the Born cross section.
\par
An interesting property of angular distributions for $\PW\gamma$ pair
production is the radiation zero, \ie a kinematical configuration
where all tree-level helicity amplitudes of the parton process are
exactly zero.  The radiation zero appears at
\begin{equation}
\cos\hat{\theta}^{\gamma}=\frac{Q_\Pu+Q_\Pd}{Q_\Pu-Q_\Pd}=\frac{1}{3},
\label{cosradzero}
\end{equation}
where $\hat{\theta}$ is the scattering angle of the photon in the
partonic CM frame, and $\Qu$ and $\Qd$ denote the charge of the up
quark and down quark, respectively. At the LHC, the radiation zero
should be observable as a dip in the distribution in the rapidity
difference between the photon and the charged lepton coming from the
$\PW$~boson \cite{Baurradzero,DeFlorianWg}. It arises from gauge
cancellations, and is characteristic for the SM. 
As a consequence,
deviations from the SM gauge structure generally tend to fill in the
dip, such that an analysis of the radiation zero provides an excellent
opportunity to probe new physics. QCD corrections generally enhance
the cross section at the radiation zero \cite{Baurradzero}.  Depending
on the process definition, the effect can be quite dramatic.  If no
jet veto is applied, the dip arising from the radiation zero is
completely washed out. If selecting only events without final-state
jets, a dip at the level of 20\% survives. 
In the bottom right plot of 
\reffi{resultsenu1}, we display our results for the distribution in
the rapidity difference between the photon and the charged lepton,
which exhibits the dip originating from the radiation zero at
$\De\eta^{l\ga}\approx 0$. The electroweak corrections hardly
influence the dip and amount to about $-2\%$. This 
is probably too small to have an observable impact on the data
analysis. The smallness of the $\Oa$ correction has again its origin
in the fact that a given bin receives the main contributions from the
low-energy domain of phase-space, where the electroweak corrections
are small.  Selecting kinematical regions characterized by larger
energies and scattering angles would increase the size of the
electroweak corrections, at the same time enhancing the radiation zero
dip, as shown in \citere{Elena}, but decrease the statistics.  

Analogous remarks apply to the electroweak corrections to the
photon-rapidity distribution shown in the bottom left of
\reffi{resultsenu1}.  The central peak in this distribution is caused
by events with small $p^\ga_{\mathrm{T}}$ and disappears if the cut
$p^{\ga,\mathrm{c}}_{\mathrm{T}}=50\GeV$ is increased. The peaks in
forward and backward direction are also present for larger
$p^{\ga,\mathrm{c}}_{\mathrm{T}}$.  They result from the peaking of
the partonic matrix element in the direction of the incoming up quark
which is additionally amplified by the boost from the CM to the
laboratory system, which goes preferably along the direction of the
valence quark.

\subsection{Comparison with virtual correction in high-energy approximation}

It is interesting to compare our results for the electroweak
corrections to hadronic $\PW\ga$ production with those obtained in
\citere{Elena}. In this latter calculation, the virtual corrections
were determined for the production subprocess only, and using the
High-Energy Approximation (HEA) of
\citeres{Stefano,StefanoHE1,StefanoHE2}. In this approximation, all
terms that vanish with $\MW/\sqrt{\hat{s}}$ at high energies are
neglected, and only the logarithmic contributions to the loop
integrals of the form $\ln^2(\hat{s}/\MW^2)$,
$\ln(\hat{s}/\MW^2)\ln(\hat{s}/\hat{x})$ and $\ln(\hat{s}/\MW^2)$ are
taken into account, where $\hat{x}=\hat{t},\hat{u}$ denotes one of the
Mandelstam variables for the production subprocess.  The use of the
HEA is justified by requiring a large transverse momentum of the
final-state photon, which results in large values for the partonic CM
energy $\sqrt{\hat{s}}$ compared to the $\PW$-boson mass, and thus to
large values for the above logarithms.
\par
In order to tune our results to those of \citere{Elena}, we generated
the amplitudes for the virtual corrections using {\tt Pole}, but we
switched off the diagrams involving the corrections to the decay
subprocess. We evaluated the so-obtained amplitudes numerically by
adapting the input to the parameter and cut values used in
\citere{Elena}. For this comparison, we moreover summed up the process
(\ref{PPnlbara}) and its charge-conjugate, \ie we considered both
$\PW^+\ga$ and $\PW^-\ga$ production.  The resulting numerical values
for the Born cross section and the finite virtual corrections are
shown in \refta{elenacs} for some values of the cut on the transverse
momentum of the final-state photon. The numbers in parentheses
represent the integration errors in the last digits.  Note that the
values shown in \refta{elenacs} do not exactly correspond to the
results presented in a similar table in \citere{Elena}. This is
because we found an error in the implementation of the
azimuthal-pseudorapidity separation (\ref{Rrec}) when comparing our
results to those of \citere{Elena}.
\begin{table}[t]
\tabcolsep 4.5pt
\bce
\begin{tabular}{|c|| c| c| c|| c| c| c|}
\hline
\hline
&\multicolumn{3}{|c||}{\citere{Elena}}&\multicolumn{3}{c|}{this work}\\
\hline
$p_{\rm T}^{\gamma,{\rm c}}({\rm GeV})$&$\sigma_{\Pp\Pp}^{(0)}({\rm
  fb})$&
$\sigma_{{\Pp\Pp,\rm finite}}^{{\rm virt}}({\rm fb})$&
$\de_{{\Pp\Pp,\rm finite}}^{{\rm virt}}$&
$\sigma^{(0)}_{\Pp\Pp}({\rm fb})$&
$\sigma_{{\Pp\Pp,\rm finite}}^{{\rm virt}}({\rm fb})$&
$\de_{{\Pp\Pp,\rm finite}}^{{\rm virt}}$\\
\hline
\hline
250&{\footnotesize $6.01~$}&{\footnotesize $ -0.292$}& {\footnotesize ~$-4.86\%$}&{\footnotesize $6.02(3)$}&{\footnotesize 
$-0.649(3)$}&{\footnotesize $-10.8\%$}\\
\hline
450&{\footnotesize $0.712$}&{\footnotesize $-9.78\times10^{-2}$}&{\footnotesize $-13.74\%$}&{\footnotesize $0.711(4)$}&
{\footnotesize $-0.1372(7)$}&{\footnotesize $-19.3\%$}\\
\hline
700&{\footnotesize $9.30\times10^{-2}$}&{\footnotesize $-2.06\times10^{-2}$}&{\footnotesize $-22.12\%$}&{\footnotesize 
$9.31(6)\times10^{-2}$}&{\footnotesize $-2.54(2)\times10^{-2}$}&{\footnotesize $-27.3\%$}\\
\hline
1000&{\footnotesize $1.25\times 10^{-2}$}&{\footnotesize $-3.72\times10^{-3}$}&{\footnotesize $-29.6\%$~}&
{\footnotesize $1.253(9)\times10^{-2}$}&{\footnotesize $-4.32(3)\times10^{-3}$}&{\footnotesize $-34.5\%$}\\
\hline
\end{tabular}
\ece
\caption{Lowest-order cross section and finite virtual
  corrections to the production subprocess for $\Pp\Pp\rightarrow
  \PW^\pm\gamma + X\rightarrow l\nu_l\gamma + X$, where $l\nu_l=l\bar\nu_l+
\bar l\nu_l$ and $l=e,\mu$.}
\label{elenacs} 
\end{table}
%
% table with HEA results
%\begin{table}[t]
%\hspace*{-0.25cm}
%\begin{tabular}{|c|| c| c| c|| c| c| c|}
%\hline
%\hline
%&\multicolumn{3}{|c||}{\citere{Elena}}&\multicolumn{3}{c|}{this work}\\
%\hline
%$p_{\rm T}^{\gamma,{\rm c}}({\rm GeV})$&$\sigma_{\Pp\Pp}^{(0)}({\rm fb})$&\multicolumn{2}{|c||}{$\sigma_{{\Pp\Pp,\rm 
%finite}}^{{\rm virt}}({\rm fb})$}&$\sigma^{(0)}_{\Pp\Pp}({\rm
%fb})$&\multicolumn{2}{|c|}{$\sigma_{{\Pp\Pp,\rm finite}}^
%{{\rm virt}}({\rm fb})$}\\
%\hline
%\hline
%250&{\footnotesize $6.01~$}&{\footnotesize $ -0.292$}& {\footnotesize ~$-4.86\%$}&{\footnotesize $6.02(3)$}&{\footnotesize 
%$-0.637(3)$}&{\footnotesize $-10.6\%$}\\
%\hline
%450&{\footnotesize $0.712$}&{\footnotesize $-9.78\times10^{-2}$}&{\footnotesize $-13.74\%$}&{\footnotesize $0.711(4)$}&
%{\footnotesize $-0.1343(7)$}&{\footnotesize $-18.9\%$}\\
%\hline
%700&{\footnotesize $9.30\times10^{-2}$}&{\footnotesize $-2.06\times10^{-2}$}&{\footnotesize $-22.12\%$}&{\footnotesize 
%$9.31(6)\times10^{-2}$}&{\footnotesize $-2.49(2)\times10^{-2}$}&{\footnotesize $-26.8\%$}\\
%\hline
%1000&{\footnotesize $1.25\times 10^{-2}$}&{\footnotesize $-3.72\times10^{-3}$}&{\footnotesize $-29.6\%$~}&
%{\footnotesize $1.253(9)\times10^{-2}$}&{\footnotesize $-4.26(3)\times10^{-3}$}&{\footnotesize $-34.0\%$}\\
%\hline
%\end{tabular}
%\caption{\label{elenacs} Lowest-order cross section and finite virtual
%  corrections to the production subprocess for $\Pp\Pp\rightarrow
%  \PW^\pm\gamma + X\rightarrow l\nu_l\gamma + X$, where $l\nu_l=l\bar\nu_l+
%\bar l\nu_l$ and $l=e,\mu$.}
%\end{table}\par
The results for the Born cross section agree within errors at the
per-mille level. Comparing our results for the finite virtual
corrections to those obtained in the HEA in \citere{Elena}, one would
expect a sizeable difference for low $p_{\rm T}^{\rm \ga,\mathrm{c}}$
values, which decreases for higher values of this cut to the level of
$1$--$2\%$.  However, the numbers in \refta{elenacs} point rather
towards a constant difference of approximately $5$--$6\%$. 

In order to understand this effect, we used the fact that {\tt Pole}
employs the FF-based package {\tt LoopTools} \cite{FF,LoopTools},
which numerically reduces all tensorial loop integrals onto a set of
scalar basis integrals. We replaced each loop integral in this basis
set by the corresponding expression in the HEA, and split the latter
into
different contributions. The total electroweak corrections obtained in
this way are smaller than those in \refta{elenacs} by at most half a per
cent. The subcontributions obtained with this procedure are shown in
\refta{elenacsvirt}.
\begin{table}
\tabcolsep 5.3pt
\bce
\begin{tabular}{|c||  c| c|| c| c|| c| c| }
%\hline
%\multicolumn{7}{|c|}{{\bf Subcontributions to \boldmath{$\sigma^{(1)}$}}}\\
\hline
\hline
$p_{\rm T}^{\gamma,\rm c}({\rm GeV})$&
$\sigma^{(1)}_{{\rm EWLog}}({\rm fb})$&
$\de^{(1)}_{{\rm EWLog}}$&
$\Delta \sigma^{(1)}_{{\rm QED}}({\rm fb})$&
$\de^{(1)}_{{\rm QED}}$&
$\sigma^{(1)}_{{\rm rem}}({\rm fb})$&
$\de^{(1)}_{{\rm rem}}$\\
\hline
\hline
250&{\footnotesize $-0.364(1)$}&{\footnotesize $-6.05\%$}&{\footnotesize $8.6(1)\times10^{-2}$}&\hspace{0.1cm}%
{\footnotesize$1.43\%$}\hspace*{0.1cm}&{\footnotesize $-0.360(2)$}&\hspace{0.1cm}{\footnotesize $-5.99\%$}%
\hspace*{0.1cm}\\
\hline
450&{\footnotesize $-0.1026(4)$}&{\footnotesize $-14.4\%$}&{\footnotesize $1.13(2)\times10^{-2}$}&{\footnotesize 
$1.59\%$}&{\footnotesize $-4.29(2)\times10^{-2}$}&{\footnotesize $-6.04\%$}\\
\hline
700&{\footnotesize $-2.11(1)\times10^{-2}$}&{\footnotesize $-22.6\%$}&{\footnotesize $1.40(3)\times10^{-3}$}&
{\footnotesize $1.50\%$}&{\footnotesize $-5.31(3)\times10^{-3}$}&{\footnotesize $-5.71\%$}\\
\hline
1000&{\footnotesize $-3.77(3)\times10^{-3}$}&{\footnotesize $-30.1\%$}&{\footnotesize $1.74(5)\times10^{-4}$}&
{\footnotesize $1.39\%$}&{\footnotesize $-6.63(5)\times10^{-4}$}&{\footnotesize $-5.29\%$}\\
\hline
\end{tabular}
\ece
\caption{Finite virtual corrections to the production subprocess for
  $\Pp\Pp\rightarrow\PW^\pm\gamma + X\rightarrow l\nu_l\gamma + X$
 in HEA split into electroweak logarithmic contributions 
 $\sigma^{(1)}_{{\rm EWLog}}$, the QED remnants 
 $\Delta \sigma^{(1)}_{{\rm QED}}$, as well as the remaining
 contributions $\sigma^{(1)}_{{\rm  rem}}$.} 
\label{elenacsvirt}
\end{table}\par
The electroweak logarithmic contributions in the second and third
column of \refta{elenacsvirt} result from the contributions to the
basic loop integrals that involve only logarithms of the form
$\ln^2(\hat{s}/\MW^2)$, $\ln(\hat{s}/\MW^2)\ln(\hat{s}/\hat{x}),$
$x=\hat{t},\hat{u}$ and $\ln(\hat{s}/\MW^2)$.  This is the approximation
used in \citere{Elena}.  As can be seen by comparing these numbers to
the results in the third and fourth column in \refta{elenacs}, the
leading logarithmic contributions differ by about $1.2\%$ for $p_{\rm
  T}^ {\ga,\mathrm{c}}=250\GeV$ and this difference decreases to the
per-mille level for higher cut values.  Thus we can reproduce the
results of \citere{Elena} to a satisfying accuracy by keeping only the
electroweak logarithms.
\par 

The QED contributions in the fourth and fifth column of
\refta{elenacsvirt} are obtained by keeping only terms involving
logarithms in the regulators for the masses of the photon and the
external light fermions, \ie logarithms of the form $\ln(\MA^2/\MW^2)$
or $\ln(m_i^2/\MW^2)$, in the loop integrals and subtracting the finite
virtual corrections as given in \refeq{virtsingLPA}. We here observe a
roughly energy-independent contribution of about $1.5\%$.  
This contribution results from the fact that the subtraction term
\refeq{virtsingLPA} involves logarithms of the form $\ln(\MA^2/s_{ij})$
and $\ln( m_i^2/s_{ij})$ while in the HEA they appear as
$\ln(\MA^2/\MW^2)$ and $\ln( m_i^2/\MW^2)$.  
\par 
Finally, the contributions listed in the sixth and seventh column of
\refta{elenacsvirt} result from all remaining terms in the HEA,
\ie logarithms depending only on ratios of kinematical variables and
constant terms in the high-energy expressions of the basic loop
integrals, as well as extra constant terms arising from the reduction
of tensor loop integral to scalar loop integrals. These remaining
terms yield a contribution of $5$--$6\%$ and thus make up the largest
part of the difference between our results and those of
\citere{Elena}.
\par 
In order to trace the origin of this large effect, we further split
the remaining contributions to the basic loop integrals as shown in
\refta{elenacsvirtconst}.
\begin{table}
\tabcolsep 4pt
\begin{center}
\begin{tabular}{|c||  c| c|| c| c|| c| c| }
%\hline
%\multicolumn{7}{|c|}{{\bf Subcontributions to \boldmath{$\sigma^{(1)}_{{\rm rem}}$}}}\\
\hline
\hline
$p_{\rm T}^{\rm c}({\rm GeV})$&
$\sigma^{(1)}_{{\rm rem, NoLog}}({\rm fb})$&
$\de^{(1)}_{{\rm rem, NoLog}}$&
$\sigma^{(1)}_{{\rm rem, Log}}({\rm fb})$&
$\de^{(1)}_{{\rm rem, Log}}$&
$\sigma^{(1)}_{{\rm rem, Log}^2}({\rm fb})$&
$\de^{(1)}_{{\rm rem, Log}^2}$\\
\hline
\hline
250&{\footnotesize $-1.55(8)\times10^{-2}$}&{\footnotesize$-0.26\%$}&{\footnotesize $-0.107(1)$}&
{\footnotesize$-1.78\%$}\hspace{0.1cm}&{\footnotesize $-0.238(2)$}&{\footnotesize $-3.95\%$}\\
\hline
450&{\footnotesize $-4.13(8)\times 10^{-3}$}&{\footnotesize $-0.58\%$}&{\footnotesize $1.139(2)\times10^{-2}$}&
{\footnotesize $-1.97\%$}&{\footnotesize $-2.48(2)\times10^{-2}$}&{\footnotesize $-3.50\%$}\\
\hline
700&{\footnotesize $-6.94(9)\times10^{-4}$}&{\footnotesize $-0.74\%$}&{\footnotesize $-1.78(2)\times10^{-3}$}&
{\footnotesize $-1.91\%$}&{\footnotesize $-2.83(2)\times10^{-3}$}&{\footnotesize $-3.05\%$}\\
\hline
1000&{\footnotesize $-1.09(1)\times10^{-4}$}&{\footnotesize $-0.87\%$}&{\footnotesize $-2.25(3)\times10^{-4}$}&
{\footnotesize $-1.79\%$}&{\footnotesize $-3.29(3)\times10^{-4}$}&{\footnotesize $-2.62\%$}\\
\hline
\end{tabular}
\end{center}
\caption{The non-leading contributions to the
  virtual corrections  to the production subprocess for
  $\Pp\Pp\rightarrow\PW^\pm\gamma + X\rightarrow l\nu_l\gamma + X$
 in HEA split up into terms $\sigma^{(1)}_{{\rm rem,
      NoLog}}$ containing no angular-dependent logarithms, into terms
  $\sigma^{(1)}_{{\rm  rem,Log}}$ depending on single
  angular-dependent logarithms as well as into terms
  $\sigma^{(1)}_{{\rm  rem, Log}^2}$ depending on angular-dependent
  logarithms squared.} 
\label{elenacsvirtconst}
\end{table}
The numbers in the fourth and fifth column comprise the contributions
linear in the purely angular-dependent logarithms, \ie logarithms of
the form $\ln(\hat{s}/\hat{x}),$ $x=\hat{t},\hat{u}$, whereas the
numbers in the sixth and seventh column correspond to the terms
containing these logarithms squared.  As can be seen from
\refta{elenacsvirtconst}, the contributions linear and quadratic in
the purely angular-dependent logarithms vary only little with
increasing energy, but contribute with about $-2\%$ and $-(3$--$4)\%$,
respectively, \ie the make up the bulk of the remaining contributions.
The large size of the contributions of the purely angular-dependent
logarithms can be explained by the enhancement of the cross section of
the considered processes in the forward and backward directions.
Since in these regions $\hat{s}/\hat{t}$ or $\hat{s}/\hat{u}$ are
small the purely angular-dependent logarithms become large. The
contributions of these logarithms are the source of comparably large
difference between the corrections in the HEA of \citere{Elena} and
the results of this paper.  On the other hand, as shown in the second
and third column of \refta{elenacsvirtconst} the non-logarithmic
remaining terms contribute to the cross section only below the
per-cent level. The increase of this contribution with rising
transverse momentum cut of the photon can be attributed to the energy
dependence of those parts of the matrix elements that have not been
evaluated in the HEA.
\par 
We thus find that, for processes dominated by contributions coming
from phase-space regions with small kinematical variables, the
accuracy of the HEA which only takes into account enhanced logarithms
of $\hat s/M^2$ may only be at the level of several per cent, unless
the enhanced angular-dependent logarithms are included as well.

\section{Summary and Conclusions}
\label{secconclusion}
We have calculated the electroweak corrections to $\PZ\gamma$ and
$\PW\gamma$ production at the LHC, with $\PZ$ and $\PW$ decaying into
leptons. For the Born cross sections and the real corrections we use
complete matrix elements.  The virtual corrections are included in the
leading-pole approximation.  For a treatment of the soft and collinear
divergences, we used the dipole 
subtraction method as well as the
phase-space-slicing technique.

We have implemented our strategy in a Mathematica package called {\tt
  Pole}, which works as an extension of the programs {\tt FeynArts3}
and {\tt FormCalc3.1} and is designed to calculate electroweak
corrections in leading-pole approximation for hadronic or partonic
processes. We have applied the program to the $\Pp\Pp\rightarrow
\nu_l\bar{\nu_l}\ga , l\bar{l}\ga ,\nu_l\bar{l}\ga$ processes at the
LHC, and tested the results extensively. As part of our checks we have
performed a comparison with the virtual corrections evaluated in the
high-energy approximation, which reveals that angular-dependent
logarithms can give effects of several per cent at high energies.

For the above-mentioned processes and typical LHC cuts, we find
electroweak corrections of the order of $-5\%$ for total cross
sections and angular distributions as, for instance, the radiation
zero in $\PW\gamma$ production.  For transverse-momentum and energy
distributions, the $\Oa$ corrections contribute up to $-20\%$ of the
lowest-order cross section. Their influence is thus well above the
systematic and statistical errors, and may be of the same size as the
QCD corrections or even larger, depending on the cut settings.
Consequently, electroweak radiative effects have to be included in the
experimental analysis when looking 
for the existence of anomalous vector-boson couplings or for the
production of new particles via hadronic production of $\PZ\gamma$ and
$\PW\gamma$.  Thus, the Monte Carlo generators for LHC should include
besides QCD corrections also the electroweak corrections.

\section*{Acknowledgements}
We thank M.~Roth for his invaluable help concerning the Monte Carlo
generator and S.~Dittmaier for carefully reading the manuscript.  
This work was supported in part by the Italian Ministero
dell'Istruzione, dell'Universit\`a e della Ricerca (MIUR) under
contract Decreto MIUR 26-01-2001 N.13.

\appendix
\def\chaptername{Appendix}
\section{Analytical results for the amplitudes}
\label{explicitamps}
In this appendix, we summarize analytical results for those parts of
the amplitudes that are compact enough to be explicitly displayed.
We present the analytical expressions for the Born amplitude in
Section \ref{secborn}, the amplitude corresponding to the
non-factorizable corrections in Section \ref{secnfac}, and the
amplitudes for the QED bremsstrahlung processes in
\refse{secbremsQED}. The calculation of the Born and bremsstrahlung
amplitudes are carried out in the framework of the Weyl--van der
Waerden formalism \cite{Weyl,vdWaerden} as presented in
\citere{Dittmaier}.

\subsection{The Lowest-order amplitudes}
\label{secborn}
\begin{figure}
\epsfig{file=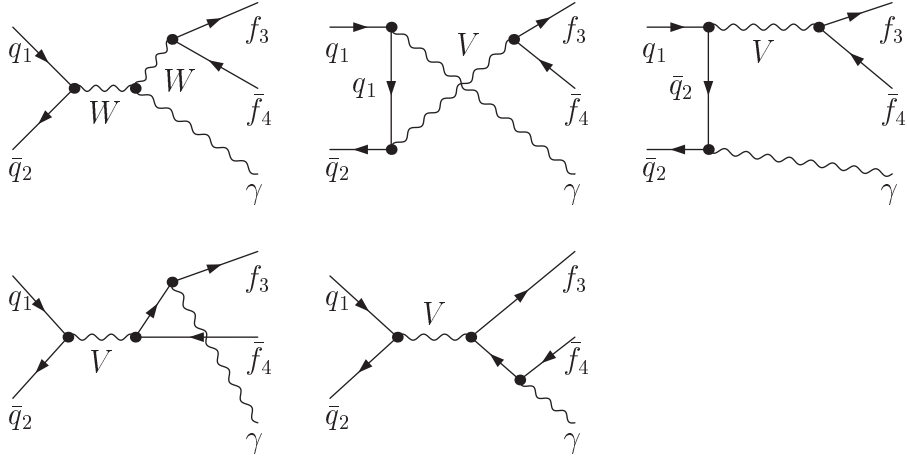}
\caption{Tree-level diagrams for $q_1 + \bar{q}_2 \rightarrow f_3 + \bar{f}_4 + \gamma$} 
\label{gendiagrams}  
\end{figure}

The helicity amplitudes of all partonic processes (\ref{qqffbara})
can be obtained from the generic set of Feynman diagrams shown in 
\reffi{gendiagrams}.
The amplitude corresponding to these diagrams can be written as
\begin{multline}
{\cal M}^{(0)\;\sigma_1 \sigma_2 \sigma_3 \sigma_4 \lambda_5}_{\;\;c_1 c_2 }(p_1,p_2,p_3,p_4,p_5)
=-2\sqrt{2} e^3\delta_{c_1c_2}\delta_{\sigma_1 -\sigma_2} \delta_{\sigma_3 -\sigma_4}\\
\sum_{V}C^{\sigma_1}_{V \bar{q}_2 q_1} C^{\sigma_3}_{V \bar{f}_3 f_4}
 {\rm B}_{V}^{\sigma_1 \sigma_3 \lambda_5}(p_1,p_2,p_3,p_4,p_5,Q_{q_1},Q_{{q}_2},Q_{f_3},Q_{{f}_4}),
\label{splitcolor}
\end{multline}
where the quantities $Q_i, i=1,\ldots 4$ denote the charges of the
external fermions (the charges of the external antifermions are given
by $-Q_i$.) and $c_1,c_2$ are the colour indices for the initial-state
quarks. The sum over $V$ runs over all SM vector
bosons. The non-vanishing gauge couplings $C^{\pm}_{V \bar{f}_a f_b} $
of a fermion $f_b$ and an anti-fermion $\bar{f}_a$ to the vector boson
$V$ are listed in \refta{gaugecouplingtable} using the following
conventions.
\begin{table}[t]
\begin{tabular}{l c l l}
$C^{\pm}_{\gamma \bar{f_2} f_1}$&$=$&$-Q_{f_1} \delta_{f_1 f_2},$&\quad$f_{1,2}=q,l$\\[2ex]
$C^{\sigma}_{Z \bar{f}_2 f_1}$&$=$&$\left\{
\begin{array}{l c l}
-\frac{\sw}{\cw} Q_{f_1}\delta_{f_1f_2},&\quad& {\rm for}\; \sigma=+\\[1ex]
\frac{I^3_{f_1}-\sw^2Q_{f_1}}{\sw \cw} \delta_{f_1 f_2},&\quad& {\rm for}\; \sigma=-
\end{array}
\right.,$&\quad$f_{1,2}=q,l$\\[4ex]
$C^{\sigma}_{W^{-} \bar{l} \nu_l}$&$=$&$\left\{
\begin{array}{l c l}
\rlap{0,}%
\hphantom{\frac{I^3_{f_1}-\sw^2Q_{f_1}}{\sw \cw} \delta_{f_1 f_2},}%
&\quad& {\rm for}\; \sigma=+\\[1ex]
\frac{1}{\sqrt{2} \sw },&& {\rm for}\; \sigma=-
\end{array}
\right.,$&\quad$l=\Pe,\mu,\tau$\\[4ex]
$C^{\sigma}_{W^{+} \bar{\nu}_l l}$&$=$&$\left\{
\begin{array}{l c l}
\rlap{0,}%
\hphantom{\frac{I^3_{f_1}-\sw^2Q_{f_1}}{\sw \cw} \delta_{f_1 f_2},}%
&\quad& {\rm for}\; \sigma=+\\[1ex]
\frac{1}{\sqrt{2} \sw },&& {\rm for}\; \sigma=-
\end{array}
\right.,$&\quad$l=\Pe,\mu,\tau$\\[4ex]
$C^{\sigma}_{W^{-} \bar{q}_2 q_1}$&$=$&$\left\{
\begin{array}{l c l}
\rlap{0,}%
\hphantom{\frac{I^3_{f_1}-\sw^2Q_{f_1}}{\sw \cw} \delta_{f_1 f_2},}%
&\quad& {\rm for}\; \sigma=+\\[1ex]
\frac{1}{\sqrt{2} \sw }V_{q_2 q_1}^{\dagger},&\quad& {\rm for}\; \sigma=-
\end{array}
\right.,$&\quad$q_{1}=\Pu,\Pc,\Pt,\quad q_{2}=\Pd,\Ps,\Pb $\\[4ex]
$C^{\sigma}_{W^{+} \bar{q}_2 q_1}$&$=$&$\left\{
\begin{array}{l c l}
\rlap{0,}%
\hphantom{\frac{I^3_{f_1}-\sw^2Q_{f_1}}{\sw \cw} \delta_{f_1 f_2},}%
&\quad& {\rm for}\; \sigma=+\\[1ex]
\frac{1}{\sqrt{2} \sw }V_{q_2 q_1},&\quad& {\rm for}\; \sigma=-
\end{array}
\right.,$&\quad$q_{1}=\Pd,\Ps,\Pb,\quad q_{2}=\Pu,\Pc,\Pt$
\end{tabular}
\caption{\label{gaugecouplingtable} Coupling constants of quarks and
  leptons to the gauge bosons $V=\gamma, Z, W^{\pm}$} 
\end{table}%
The particle indices $l$ ($q$) denote leptons (quarks). The charges
and the third component of the isospin of the fermion $f$ are denoted
by $Q_f$ and $I^3_f$, respectively. The quantities $\sw$ and $\cw$ are the sine and the cosine of the weak mixing angle, whereas
$\alpha$ is the fine-structure constant. Finally, $\delta_{ij}$ and
$V_{ij}$ denote the Kronecker delta and the quark mixing matrix. Note,
that the elementary charge $e$ has been extracted from the
couplings.
\par 
As discussed in \refse{subsectiondecays}, the amplitude can be split
into a resonant and  a non-resonant part. This splitting
can be done on the level of the generic functions
\begin{equation}
{\rm B}_{V}^{\sigma_1 \sigma_3 \lambda_5}={\rm B}_{V}^{{\rm res},\sigma_1 \sigma_3 \lambda_5}+{\rm B}_{V}^{{\rm nres},\sigma_1 \sigma_3 \lambda_5}.
\end{equation}
The resonant part comprises all contributions from diagrams containing
the resonant propagator of the decaying particle $V$, and thus
corresponds to the amplitude originating from the diagrams in the
first row of \reffi{gendiagrams}. Applying the Weyl--van der
Waerden formalism as presented in \citere{Dittmaier} to these
diagrams, the generic function for all helicities positive is given by
\begin{multline}
{\rm B}_{V}^{{\rm res},+ + +}(p_1,p_2,p_3,p_4,p_5,Q_1,Q_2,Q_3,Q_4)
=-P_V(p_3+p_4)\frac{\pr{p_1}{p_4}^2}{\pr{p_4}{p_5}}\\
\times\left\{\frac{Q_1\cpr{p_3}{p_2}}{\pr{p_1}{p_5}}+\frac{Q_2\cpr{p_3}{p_1}}{\pr{p_2}{p_5}}-(Q_2-Q_1)P_V(p_1+p_2)\cpr{p_3}{p_5}\cpr{p_2}{p_1}\right\}.
\label{bornfac}
\end{multline}
The non-resonant part of the Born amplitude, on the other hand,
receives contributions from the remaining diagrams in the second row
of \reffi{gendiagrams}. Expressed in Weyl-spinor products, the
generic function for all helicities positive reads
\begin{equation}
{\rm B}_{V}^{{\rm nres},+ + +}(p_1,p_2,p_3,p_4,p_5,Q_1,Q_2,Q_3,Q_4)
=P_V(p_1+p_2)\frac{Q_3 \pr{p_1}{p_4}^2 \cpr{p_2}{p_1}}{\pr{p_3}{p_5} \pr{p_4}{p_5}}.
\label{bornnfac}
\end{equation}
Here $M_V$ is the mass and $\Gamma_V$ the decay width of the vector
boson $V$, and the corresponding propagators are denoted as
\begin{equation}
P_V(k)=\frac{1}{k^2-M_V^2-\ri M_V\Gamma_V}.
\end{equation}
The Weyl-spinor products are defined by
\beq
\langle pq\rangle
=2\sqrt{p_0 q_0} \,\Biggl[
{\mathrm{e}}^{-\ri\phi_p}\cos\frac{\theta_p}{2}\sin\frac{\theta_q}{2}
-{\mathrm{e}}^{-\ri\phi_q}\cos\frac{\theta_q}{2}\sin\frac{\theta_p}{2}
\Biggr],
\eeq
where $\theta_p$, $\theta_q$ and $\phi_p$, $\phi_q$ are the polar and
azimuthal angles, respectively, of the corresponding light-like
4-momenta
\beqar
p^\mu&=&p_0(1,\sin\theta_p\cos\phi_p,\sin\theta_p\sin\phi_p,\cos\theta_p),\nl
q^\mu&=&q_0(1,\sin\theta_q\cos\phi_q,\sin\theta_q\sin\phi_q,\cos\theta_q).
\eeqar

The amplitudes for neutral current reactions can be extracted from
(\ref{bornfac}) and (\ref{bornnfac}) by simply setting $(Q_2-Q_1)=0$.
Inserting (\ref{bornfac}) and (\ref{bornnfac}) in (\ref{splitcolor})
and adding the resulting expressions yields the full Born amplitude.
While this is gauge invariant, the resonant and non-resonant parts as
defined above are gauge-dependent, unless applying the LPA.
To this end, one has to set the finite width in the $s$-channel
propagator $P_V(p_1+p_2)$ in the resonant amplitude (\ref{bornfac}) to
zero and to insert the resulting expression in (\ref{splitcolor}),
where the spinor and scalar products have to be evaluated using the
on-shell projected momenta, except for in the resonant propagator
$P_V(p_3+p_4)$. Note that we set the gauge spinor for the external
photon to $g=p_4$ to obtain the above results, which is why the
contribution of the diagram with the photon coupling to the external
fermion $f_4$ vanishes.
\par 
Dropping the superscript for resonant and non-resonant contributions
in what follows, the other helicity combinations can be obtained as
\begin{align}
{\rm B}_{V}^{- + +}(p_1,p_2,p_3,&\,p_4,p_5,Q_1,Q_2,Q_3,Q_4)\nl
&={\rm B}_{V}^{+ + +}(p_2,p_1,p_3,p_4,p_5,-Q_2,-Q_1,Q_3,Q_4),\nl
{\rm B}_{V}^{+ - +}(p_1,p_2,p_3,&\,p_4,p_5,Q_1,Q_2,Q_3,Q_4)\nl
&={\rm B}_{V}^{+ + +}(p_1,p_2,p_4,p_3,p_5,Q_1,Q_2,-Q_4,-Q_3),\nl
{\rm B}_{V}^{- - +}(p_1,p_2,p_3,&\,p_4,p_5,Q_1,Q_2,Q_3,Q_4)\nl
&={\rm B}_{V}^{+ + +}(p_2,p_1,p_4,p_3,p_5,-Q_2,-Q_1,-Q_4,-Q_3),
\end{align}
as well as
\begin{multline}
{\rm B}_{V}^{\sigma_1 \sigma_3 -}(p_1,p_2,p_3,p_4,p_5,Q_1,Q_2,Q_3,Q_4)\\=
\left[{\rm B}_{V}^{-\sigma_1 -\sigma_3 +}(p_1,p_2,p_3,p_4,p_5,Q_1,Q_2,Q_3,Q_4)\right]^{\star}\Bigg\vert_{P_V^\star \rightarrow P_V}.
\label{otherhelicities}
\end{multline}

\subsection{The non-factorizable corrections}
\label{secnfac}

A general discussion for the photonic non-factorizable corrections for
a process involving massless external particles and an arbitrary
number of resonances can be found in \citere{Kaiser}.  Applying the
general results displayed there to our processes, the correction
factor results in
\begin{multline}
\delta^{{\rm virt}}_{{\rm nfac}}(\Phi^{\rm
  osh}_0,(p_3+p_4)^2-M_V^2)\\=\sum_{i=1,2}\sum_{j=3,4}\tau_i\tau_jQ_iQ_j\frac{\alpha}{\pi}\Re\left\{\Delta^V(p_i^{\rm osh};(p_3+p_4),p_j^{\rm osh})\right\},\quad V=\PW,\PZ 
\label{deltanf}
\end{multline}
with 
\begin{multline}
\Delta^V(p_i^{\rm osh};(p_3+p_4),p_j^{\rm
  osh})\\=2\ln\left(\frac{\MA M_V}{ M_V^2-\ri M_V \Gamma_V-(p_3+p_4)^2}\right)\left[\ln\left(\frac{\hat{t}}{t}\right)-1\right]-2-{\rm Li}_2\left(1-\frac{\hat{t}}{t}\right).
\end{multline}
The kinematical variables $\hat{t}$ and $t$ are defined by 
\begin{equation}
t=(p_i^{\rm osh}-p_j^{\rm osh})^2, \qquad \hat{t}=(p_i^{\rm
  osh}-p_3^{\rm osh}-p_4^{\rm osh})^2-M_V^2, 
\end{equation}
and ${\rm Li}_2$ denotes the dilogarithm (\ref{dilogdef}). 

\subsection{The real QED corrections}
\label{secbremsQED}
The determination of the helicity amplitudes for the real QED
corrections can be performed in the same way as described
in Section \ref{secborn} for the case of the Born amplitudes. The
calculation is much more involved, though, since the set of generic
diagrams comprises here 31 instead of five diagrams. These diagrams
can be obtained from the ones shown in \reffi{gendiagrams} in the
following way. A total number of thirty diagrams is obtained by
attaching the additional real photon to each fermion line and to each
massive vector-boson line in the Born graphs shown in 
\reffi{gendiagrams}. The one remaining diagram results from both photons
coupling to the intermediate vector boson via a $VV\gamma \gamma$
coupling. Apart from the increased number of diagrams, one has to
calculate the generic amplitude for two different polarization
combinations, since a final state with both photons having the same
helicity is not related by any discrete symmetry to a final state with
two photons of opposite helicity.\par Adopting the conventions of
Section \ref{secborn}, the generic amplitude of the QED bremsstrahlung
process reads
\begin{multline}
{\cal M}^{\;\sigma_1 \sigma_2 \sigma_3 \sigma_4 \lambda_5 \lambda_6}_{{\rm real,QED}\;\; c_1 c_2 c_3 c_4}(p_1,p_2,p_3,p_4,p_5,p_6)
=-4 e^4 \delta_{c_1c_2}\delta_{\sigma_1 -\sigma_2} \delta_{\sigma_3 -\sigma_4}\\
\sum_{V}C^{\sigma_1}_{V \bar{q}_2 q_1} C^{\sigma_3}_{V \bar{f}_3 f_4}
{\rm B}_{V \gamma}^{\sigma_1 \sigma_3 \lambda_5 \lambda_6}(p_1,p_2,p_3,p_4,p_5,p_6,Q_{q_1},Q_{{q}_2},Q_{f_3},Q_{{f}_4}),
\end{multline}
where the sum over $V$ is again taken over all SM vector bosons. The
couplings of the gauge boson $V$ to the fermion $f_a$ and the
anti-fermion $\bar{f}_b$ is again denoted by $C^{\sigma_a}_{V
  \bar{f}_b f_a}$ and can be read off \refta{gaugecouplingtable}.
The generic function for all helicities positive is obtained as
\begin{eqnarray}
\lefteqn{{\rm B}_{V \gamma}^{+ + + +}(p_1,p_2,p_3,p_4,p_5,p_6,Q_1,Q_2,Q_3,Q_4)=\frac{\pr{p_4}{p_1}^2}{\pr{p_4}{p_5}\pr{p_4}{p_6}}}\qquad\nl
&&{}\times\Bigg\{
P_V(p_3+p_4)
\Bigg[
\frac{-Q_1^2 \cpr{p_3}{p_2} \pr{p_4}{p_1}}{ \pr{p_1}{p_5}\pr{p_1}{p_6}}
+\frac{Q_1 Q_2 \spr{p_4}{p_2}{p_5}{p_3}}{\pr{p_2}{p_5}\pr{p_1}{p_6}}\nl
&&\qquad{}
+\frac{Q_1Q_2 \spr{p_4}{p_2}{p_6}{p_3}}{\pr{p_1}{p_5}\pr{p_2}{p_6}}
+\frac{Q_2^2\cpr{p_1}{p_3}\pr{p_2}{p_4}}{\pr{p_2}{p_5}\pr{p_2}{p_6}}
\Bigg]\nl
&&{}
+P_V(p_3+p_4+p_5)
\Bigg[
\frac{Q_3}{\pr{p_3}{p_5}}+P_V(p_3+p_4)(Q_2-Q_1)\cpr{p_5}{p_3}
\Bigg]\nl
&&\qquad{}\times
\Bigg[
\frac{Q_1\spr{p_4}{p_1}{p_6}{p_2}}{\pr{p_1}{p_6}}
+\frac{Q_2\spr{p_4}{p_2}{p_6}{p_1}}{\pr{p_2}{p_6}}
\nl&&\qquad\qquad{}
-P_V(p_1+p_2)(Q_2-Q_1)\cpr{p_1}{p_2}\spr{p_4}{p_2}{p_1}{p_6}
\Bigg]\nl
&&{}
+P_V(p_3+p_4+p_6)
\Bigg[
\frac{Q_3}{\pr{p_3}{p_6}}+P_V(p_3+p_4)(Q_2-Q_1)\cpr{p_6}{p_3}
\Bigg]\nl
&&\qquad{}\times
\Bigg[
\frac{Q_1\spr{p_4}{p_1}{p_5}{p_2}}{\pr{p_1}{p_5}}
+\frac{Q_2\spr{p_4}{p_2}{p_5}{p_1}}{\pr{p_2}{p_5}}
\nl&&\qquad\qquad{}
-P_V(p_1+p_2)(Q_2-Q_1)\cpr{p_1}{p_2}\spr{p_4}{p_2}{p_1}{p_5}\Bigg]\nl
&&{}
-P_V(p_1+p_2)
\frac{Q_3^2}{\pr{p_3}{p_5}\pr{p_3}{p_6}}\cpr{p_1}{p_2}\pr{p_3}{p_4}
\Bigg\},
\end{eqnarray}
where in the above expressions, a product like $\spr{p_1}{p_4}{p_6}{p_5}$ is defined by
\begin{equation}
\spr{a}{b}{c}{d}=\pr{a}{b}\cpr{d}{b}+\pr{a}{c}\cpr{d}{c}.
\end{equation}
The corresponding function for helicity $\lambda_5$ negative reads
\begin{eqnarray}
\lefteqn{{\rm B}_{V\gamma}^{+ + - +}(p_1,p_2,p_3,p_4,p_5,p_6,Q_1,Q_2,Q_3,Q_4)
=\frac{1}{\pr{p_4}{p_6} \cpr{p_3}{p_5}}}\qquad\nl
&&{}\times
\Bigg\{
P_V(p_3+p_4)
\Bigg[
\frac{Q_1^2}{\cpr{p_1}{p_5}\pr{p_1}{p_6}} \frac{\cpr{p_3}{p_2}}{(p_1+p_5+p_6)^2}
%\nl&&\qquad\qquad{}
%\times
\bigg(\spr{p_4}{p_1}{p_5}{p_3}\spr{p_4}{p_1}{p_5}{p_6}\pr{p_1}{p_6}
\nl&&\qquad\qquad\qquad{}
+\pr{p_1}{p_4}\cpr{p_1}{p_5}\pr{p_1}{p_5}\pr{p_2}{p_4}\cpr{p_3}{p_2}\bigg)
\nl&&\qquad{}
+\frac{Q_2^2}{\cpr{p_2}{p_5}\pr{p_2}{p_6}}\frac{\pr{p_4}{p_1}}{(p_2+p_5+p_6)^2}
%\nl&&\qquad\qquad{}\times
\bigg(\spr{p_5}{p_2}{p_6}{p_3}\spr{p_4}{p_2}{p_6}{p_3}\cpr{p_2}{p_5}
\nl&&\qquad\qquad\qquad{}
+\cpr{p_1}{p_3}\cpr{p_2}{p_6}\pr{p_2}{p_6}\cpr{p_2}{p_3}\pr{p_4}{p_1}\bigg)
\nl&&\qquad{}
+\frac{Q_1 Q_2}{\cpr{p_2}{p_5}\pr{p_1}{p_6}}\pr{p_4}{p_1}^2 \left(\cpr{p_2}{p_3}\right)^2
-\frac{Q_1Q_2}{\cpr{p_1}{p_5}\pr{p_2}{p_6}}\spr{p_4}{p_1}{p_5}{p_3}^2\Bigg]
%%%%%%%%%%%%%%%%%%%%%%%%%%%%%%%%%%%%%%%%%%%%%%%%%%
\nl&&{}
+P_V(p_3+p_4+p_5)\spr{p_1}{p_4}{p_5}{p_3}
\Bigg[
\frac{Q_4}{\cpr{p_4}{p_5}}-P_V(p_3+p_4)(Q_2-Q_1)\pr{p_5}{p_4}
\Bigg]
\nl&&\qquad{}
\times\Bigg[
\frac{Q_1}{\pr{p_1}{p_6}}\pr{p_4}{p_1}\cpr{p_3}{p_2}
-\frac{Q_2}{\pr{p_2}{p_6}}\spr{p_4}{p_2}{p_6}{p_3}
\nl&&\qquad\qquad{}
+P_V(p_1+p_2)(Q_2-Q_1)\bigg(\cpr{p_6}{p_2}\cpr{p_6}{p_3}\pr{p_6}{p_4}+\spr{p_4}{p_2}{p_1}{p_6}\cpr{p_3}{p_2}\bigg)\Bigg]
%%%%%%%%%%%%%%%%%%%%%%%%%%%%%%%%%%%%%%%%%%%%%%%%%%
\nl&&{}
+P_V(p_3+p_4+p_6)\spr{p_4}{p_3}{p_6}{p_2}
\Bigg[
\frac{Q_3}{\pr{p_3}{p_6}}+P_V(p_3+p_4)(Q_2-Q_1)\cpr{p_6}{p_3}
\Bigg]
\nl&&\qquad{}
\times\Bigg[
\frac{Q_1}{\cpr{p_1}{p_5}}\spr{p_4}{p_2}{p_6}{p_3}
+\frac{Q_2}{\cpr{p_2}{p_5}}\pr{p_4}{p_1}\cpr{p_3}{p_2}
\nl&&\qquad\qquad{}
-P_V(p_1+p_2)(Q_2-Q_1)\bigg(\pr{p_5}{p_1}\pr{p_5}{p_4}\cpr{p_5}{p_3}+\spr{p_5}{p_2}{p_1}{p_3}\pr{p_4}{p_1}\bigg)
\Bigg]
%%%%%%%%%%%%%%%%%%%%%%%%%%%%%%%%%%%%%%%%%%%%%%%%%%
\nl&&{}
+P_V(p_1+p_2)
\Bigg[P_V(p_3+p_4)(
Q_2-Q_1)^2\pr{p_1}{p_4}\cpr{p_2}{p_3}\pr{p_4}{p_5}\cpr{p_6}{p_3}
\nl&&\qquad{}
+\frac{Q_3^2}{\pr{p_3}{p_6}}\frac{\pr{p_6}{p_4}\cpr{p_3}{p_6}\pr{p_4}{p_1}}{(p_3+p_5+p_6)^2}\spr{p_5}{p_3}{p_6}{p_2}
\nl&&\qquad{}
+\frac{Q_3 Q_4}{\pr{p_3}{p_6}\cpr{p_4}{p_5}}\spr{p_1}{p_4}{p_5}{p_3}\spr{p_4}{p_3}{p_6}{p_2}
\nl&&\qquad{}
-\frac{Q^2_4}{\cpr{p_4}{p_5}}
\frac{\cpr{p_5}{p_3}\cpr{p_3}{p_2}\pr{p_5}{p_4}}{(p_4+p_5+p_6)^2 }\spr{p_1}{p_4}{p_5}{p_6}\Bigg]\Bigg\}.
\end{eqnarray}
The other helicity combinations can be calculated by
\begin{align}
{\rm B}_{V\gamma}^{- + \lambda_5 +}(p_1,p_2,p_3,&\,p_4,p_5,p_6,Q_1,Q_2,Q_3,Q_4)\nonumber\\&={\rm B}_{V\gamma}^{+ + \lambda_5 +}(p_2,p_1,p_3,p_4,p_5,p_6,-Q_2,-Q_1,Q_3,Q_4),\nonumber\\
{\rm B}_{V\gamma}^{+ - \lambda_5 +}(p_1,p_2,p_3,&\,p_4,p_5,p_6,Q_1,Q_2,Q_3,Q_4)\nonumber\\&={\rm B}_{V\gamma}^{+ + \lambda_5 +}(p_1,p_2,p_4,p_3,p_5,p_6,Q_1,Q_2,-Q_4,-Q_3),\nonumber\\
{\rm B}_{V\gamma}^{- - \lambda_5 +}(p_1,p_2,p_3,&\,p_4,p_5,p_6,Q_1,Q_2,Q_3,Q_4)\nonumber\\&={\rm B}_{V\gamma}^{+ + \lambda_5 +}(p_2,p_1,p_4,p_3,p_5,p_6,-Q_2,-Q_1,-Q_4,-Q_3),
\end{align}
and
\begin{multline}
{\rm B}_{V\gamma}^{\sigma_1 \sigma_3 \lambda_5 -}(p_1,p_2,p_3,p_4,p_5,p_6,Q_1,Q_2,Q_3,Q_4)\\=\left[{\rm B}_{V\gamma}^{-\sigma_1 -\sigma_3 -\lambda_5 +}(p_1,p_2,p_3,p_4,p_5,p_6,Q_1,Q_2,Q_3,Q_4)\right]^{\star}\Bigg\vert_{P^{\star}_V \rightarrow P_V}.
\end{multline}

\section{Explicit form of the (Lorentz-invariant) on-shell projection}
\label{app-projection}

In order to be able to evaluate the virtual corrections to
the processes 
%\refeq{PPnnbara}-\refeq{PPnlbara} 
\refeq{qqVVffbara} 
in LPA, an on-shell projection has
to be specified which maps a general set of momenta $\Phi_0$ on a set
of momenta $\Phi_0^{\rm osh}$ such that $(p^{\rm osh}_3+p^{\rm
  osh}_4)^2=M_V^2$, where $M_V$ is the mass of the decaying vector
boson $V=\PW,\PZ$. Such a projection is given by
\begin{eqnarray}
\tilde{p}_3=y_1p_3,\qquad \tilde{p}_5=y_2p_5,\qquad \tilde{p}_4=p_1+p_2-\tilde{p}_3-\tilde{p}_5,
\end{eqnarray}
where scaling variables $x$ and $z$ are obtained from the mass-shell
conditions for the resonance and for the momentum $\tilde{p}_4$ as
\begin{equation}
y_2=\frac{\hat{s}-M_V^2}{2(p_1+p_2)p_5},\qquad y_1=\frac{M_V^2}{2(p_1+p_2-zp_5)p_3}.
\end{equation}
This on-shell projection only involves Lorentz-invariant quantities.

\let\bf\relax

\bibliography{bibliography}

\end{document}